\documentclass[a4paper,11pt]{article}
\pdfoutput=1
\usepackage{jcappub}
\usepackage{amsmath}
\usepackage[T1]{fontenc}
\usepackage{bm}
\usepackage{todonotes}
\usepackage{longtable}
\usepackage{boxhandler}
\usepackage[hang,flushmargin]{footmisc}
\usepackage{caption}
\usepackage{multirow}
\usepackage{hyperref}

\newcommand{\mnras}{MNRAS}
\newcommand{\jcap}{JCAP}
\newcommand{\aap}{Astron. Astrophys.}
\newcommand{\apjl}{Astrophys. J. Lett.}

\title{Dark matter local density determination: recent observations and future prospects}
\author[a]{Pablo F.~de~Salas,}
\emailAdd{pablo.fernandez@fysik.su.se}

\affiliation[a]{The  Oskar  Klein  Centre  for  Cosmoparticle  Physics, Department  of Physics, Stockholm  University, AlbaNova, Stockholm SE-106 91, Sweden}

\author[b]{A. Widmark}
\emailAdd{axel.widmark@nbi.ku.dk}
\affiliation[b]{Dark Cosmology Centre, Niels Bohr Institute, University of Copenhagen, Jagtvej 128, 2200 Copenhagen N, Denmark}

\newcommand{\rhoDMlocal}{\ensuremath{\rho_{\mathrm{DM,\odot}}}}

\definecolor{darkGreen}{rgb}{0.0, 0.5, 0.0}

\abstract{This report summarises progress made in estimating the local density of dark matter ($\rho_{\mathrm{DM,\odot}}$), a quantity that is especially important for dark matter direct detection experiments. 
We outline and compare the most common methods to estimate $\rho_{\mathrm{DM,\odot}}$ and the results from recent studies, including those that have benefited from the observations of the ESA/Gaia satellite.
The result of most local analyses coincide within a range of $\rho_{\mathrm{DM,\odot}} \simeq \text{0.4--0.6}\,\mathrm{GeV/cm^3} = \text{0.011--0.016}\,\mathrm{M_\odot / pc^3}$, while a slightly lower range of $\rho_{\mathrm{DM,\odot}} \simeq \text{0.3--0.5}\,\mathrm{GeV/cm^3} = \text{0.008--0.013}\,\mathrm{M_\odot / pc^3}$ is preferred by most global studies.
In light of recent discoveries, we discuss the importance of going beyond the approximations of what we define as the Ideal Galaxy (a steady-state Galaxy with axisymmetric shape and a mirror symmetry across the mid-plane) in order to improve the precision of $\rho_{\mathrm{DM,\odot}}$ measurements. In particular, we review the growing evidence for local disequilibrium and broken symmetries in the present configuration of the Milky Way, as well as uncertainties associated with the Galactic distribution of baryons.
Finally, we comment on new ideas that have been proposed to further constrain the value of $\rho_{\mathrm{DM,\odot}}$, 
most of which would benefit from Gaia's final data release.
}

\begin{document}
\maketitle

\section{Introduction}\label{sec:Intro}
An extensive variety of gravitational probes indicate that something is not entirely known about the matter content of the Universe, showing that baryonic matter can only constitute a small fraction of the total mass~\cite{Bertone:2016nfn}. 
The missing matter has been noticed in a wide range of distances, from sub-galactic to cosmological scales \cite{Aghanim:2018eyx}, and all observations are consistent with particle dark matter \cite{Bergstrom:2000pn,Bertone:2004pz,Roszkowski:2017nbc}. Subscribing to this interpretation, we still do not know what type of particle, or set of particles, dark matter is made of; however, it is well constrained to a non-relativistic and non-collisional behaviour for most of cosmological history~\cite{Aghanim:2018eyx}.

In the hope that the elusive dark matter interacts with standard model particles via some force other than gravity, there are three additional ways to study its properties: via the production of dark matter particles at colliders (creation) \cite{Goodman:2010ku}; via the annihilation or decay of dark matter particles into detectable standard model particles (indirect detection) \cite{Klasen:2015uma}; or via the interaction of dark matter particles with known matter, such as a scattering target of heavy nuclei placed deep underground (direct detection) \cite{Undagoitia:2015gya}. 

Direct detection experiments look for collision events inside a detector by measuring the recoil energies of their detector medium. For a spin independent interaction between a dark matter particle of mass $m_\mathrm{DM}$ and a detector nucleus of mass $m_A$, the differential recoil rate can be expressed as
\begin{equation}\label{eq:differential-recoil-rate-SI}
\frac{\mathrm{d}R}{\mathrm{d}E_R} = \frac{\rhoDMlocal}{m_{\mathrm{DM}} m_A} \int_{v > v_{\mathrm{min}}} \mathrm{d}^3 \mathbf{v}\, v \,f(\mathbf{v}) \, \frac{\mathrm{d}\sigma_{\rm SI}}{\mathrm{d}E_R},
\end{equation}
where $v_{\rm min}$ is the minimum velocity needed to produce a nuclear recoil of energy $E_R$, $\sigma_{\rm SI}$ is the cross section, $f(\mathbf{v})$ the local velocity distribution of dark matter particles, and $\rhoDMlocal$ is the local dark matter density.
We see in eq.~\eqref{eq:differential-recoil-rate-SI} that estimating $\sigma_{\rm SI}$ from the recoil rate is contingent on sufficiently precise knowledge of $f(\mathbf{v})$ and $\rhoDMlocal$.
Moreover, $\sigma_{\rm SI}$ is degenerate with $\rhoDMlocal$, which complicates the interpretation of direct detection exclusion limits owing to the lack of a robust estimate for $\rhoDMlocal$---in section \ref{sec:new-ideas} we discuss a recent paper that suggested how to effectively break this degeneracy.
The velocity distribution is commonly assumed to be that of the \emph{standard halo model}: an isotropic, isothermal sphere with a Maxwellian distribution \cite{Green:2011bv}, although this is likely not the case according to the results from $N$-body simulations (e.g., \cite{Necib:2018iwb,Necib:2018igl,Bozorgnia:2018pfa}) and what we are learning about the history and present configuration of our own galaxy (see section \ref{sec:breaking-of-Ideal-Galaxy}). 

The local dark matter density is also an important ancillary parameter when constraining other Galactic properties. In combination with other measurements, a precise value of $\rhoDMlocal$ can be used to characterise the shape of the Galactic halo \cite{Read:2014qva,Bland-Hawthorn:1602.07702}. Furthermore, it can inform us of the possibility of a dark disc, co-planar with the stellar disc, formed either from collisionless dark matter through the accretion of satellites \cite{Read:2008fh,Purcell:2009yp} or from a dark matter sub-component with strong dissipative self-interactions \cite{Fan:2013tia,Fan:2013yva}.

Local mass measurements date back almost a century, since the works of J. C. Kapteyn \cite{Kapteyn:1922zz} and J. H. Jeans \cite{Jeans_1922} in 1922. The past few years have seen an increase in the number of publications aiming at constraining $\rhoDMlocal$, and the various techniques are getting increasingly sophisticated as the knowledge about our Galaxy improves. Yet, a clear consensus for the value of $\rhoDMlocal$ is still missing.

In this report we review the most recent efforts to estimate $\rhoDMlocal$.
To begin with, in section~\ref{sec:Ideal-Galaxy} we describe the properties of the Ideal Galaxy, which we define as a galaxy in a dynamical steady state with a matter distribution that is axisymmetric and also has a mirror symmetry with respect to the Galactic plane.
We use this definition for the Ideal Galaxy because these assumptions are very common in studies focused on the determination of $\rhoDMlocal$.
However, the mirror symmetry over the Galactic plane has been relaxed in some recent studies.
In section~\ref{sec:methods} we describe different methods that have been used to obtain the most recent $\rhoDMlocal$ estimates. These recent estimates are listed in section~\ref{sec:rhoDM-estimates}. 
A discussion on the differences in the estimates is presented in section~\ref{sec:breaking-of-Ideal-Galaxy}, where we also comment on the importance of common assumptions in the interpretation of the results, especially in view of what recent observations tell us about how much the Milky Way deviates from the Ideal Galaxy.
In section~\ref{sec:new-ideas} we comment about future prospects and new ideas to determine $\rhoDMlocal$. Finally, we conclude the report in section~\ref{sec:conclusions}.

We quote all $\rhoDMlocal$ values in units of $\mathrm{GeV/cm^3}$, where a conversion factor of $1\, \mathrm{M_\odot / pc^3} = 37.5\,\mathrm{GeV/cm^3}$ is applied to transform the results from those references giving $\rhoDMlocal$ in $\mathrm{M_\odot / pc^3}$ units.
Furthermore, any uncertainties quoted in this report correspond to $1\sigma$, unless otherwise stated.

%%%%%%%%%%%%%%%%%%%%%%%%%%%%%%%%%%%%%%%%%%%%%%%%%%%%%%%%%%%%%%%%%%%%%%
%%%%%%%%%%%%%%%%%%%%%%%%%%%%%%%%%%%%%%%%%%%%%%%%%%%%%%%%%%%%%%%%%%%%%%
\section{The Ideal Galaxy}\label{sec:Ideal-Galaxy}

Galaxies are complicated objects that can be the result of a rich history of many separate mass agglomerations, and their formation is not yet fully understood \cite{BinneyTremaine:book}. Spiral galaxies like the Milky Way contain billions of stars, the majority of which are distributed close to the Galactic plane forming a disc, while a large proportion of stars also form a bulge at the centre of the Galaxy, and a smaller amount---mostly old stars---are distributed in the shape of a stellar halo. Although the stellar content of the Milky Way sums up to $(5\pm 1)\cdot 10^{10} \,\mathrm{M_\odot}$, the Galactic virial mass is estimated to be close to $10^{12}\,\mathrm{M_\odot}$, implying that most of the mass of the Galaxy is enclosed in a dark matter halo (for a review on Galactic properties we refer the interested reader to ref.~\cite{Bland-Hawthorn:1602.07702}).

Throughout this report, we will refer to the Ideal Galaxy, representing the idealised and probably somewhat simplistic model of a spiral galaxy like the Milky Way. 
Our definition of the Ideal Galaxy includes the following three assumption: 
\begin{itemize}
    \item the Ideal Galaxy is in a \emph{steady state}, meaning that the matter density and phase-space distribution are static;
    \item the Ideal Galaxy is \emph{axisymmetric};
    \item the Ideal Galaxy has a \emph{mirror symmetry} across the Galactic plane.
\end{itemize}
These three assumptions are useful and simplifying, in fact often necessary, when performing dynamical mass measurements. They are all approximate truths, but certainly not complete truths. The question remains to what extent these assumptions will bias dynamical mass measurements, and how they can be diagnosed, quantified, and corrected for.

Given the size of the Milky Way, its stars typically need many millions of years to complete an orbit. However, this time is short with respect to the several-gigayear process of galaxy formation and age of the Milky Way. With this in mind, and since observations indicate that the last important Galactic merger happened a few gigayears ago \cite{Belokurov:1802.03414,Helmi:1806.06038}, it is not unreasonable to assume that our Galaxy has had enough time to reach a steady state that holds at present time. However, the Milky Way does definitely exhibit some time-varying dynamical structures, both on small and large scales \cite{Antoja_2018,Khoperskov:1811.09205,Binney:1807.09819}. 

In the same manner, the symmetries of the Ideal Galaxy capture the general shape of the Milky Way, although deviations with respect to these symmetries are clearly present, at both small and large spatial scales.
Current studies indicate that local axisymmetry is a viable assumption when estimating $\rhoDMlocal$, but the assumption of a mirror symmetry can lead to large biases in local analyses (see, e.g., \cite{Guo:2020rcv,Salomon:2020eer}). Out of the three assumptions included in our definition of the Ideal Galaxy, relaxing that of a mirror symmetry is the most prevalent in recent studies.

Several parametric mass models of the Milky Way, where the Galaxy has been treated as the Ideal Galaxy, can be found in the literature (e.g., \cite{Bovy:2014vfa,McMillan:1608.00971}).
Those models are very useful when looking for a simplified description of the Galaxy.
However, given that the Milky Way is not actually Ideal, such a Galactic mass model can not fit every single observation, and a compromise has to be reached, regarding the number of assumptions, in each specific analysis. Deviations from the Ideal Galaxy will be discussed from the point of view of both Galactic observations and $\rhoDMlocal$ estimates in section~\ref{sec:breaking-of-Ideal-Galaxy}.

%%%%%%%%%%%%%%%%%%%%%%%%%%%%%%%%%%%%%%%%%%%%%%%%%%%%%%%%%%%%%%%%%%%%%%
%%%%%%%%%%%%%%%%%%%%%%%%%%%%%%%%%%%%%%%%%%%%%%%%%%%%%%%%%%%%%%%%%%%%%%
\section{Methods for estimating $\rhoDMlocal$}\label{sec:methods}

In this section we present methods for estimating the local dark matter density, based on different analysis techniques and observations. 
Furthermore, tracers can be selected from a wide variety of astrophysical objects: individual stars from a given stellar population are a common choice because they perform well for almost any typical method; gas clouds close to the Galactic plane are interesting tracers for studies based on circular velocities; globular clusters could be used if $\rhoDMlocal$ were to be obtained from a fit to the shape of the Galactic halo, and streams could provide a $\rhoDMlocal$ estimate both from constraining the halo of the Galaxy and from setting bounds to the Galactic disc surface density.

The most common methods to estimate $\rhoDMlocal$ include the analysis of stellar kinematics in a local volume near the Sun (section \ref{sec:method:vertical-Jeans}); as well as the analysis of the circular velocity curve of the Galaxy (section~\ref{sec:method:Vc-fitting}).
Other approaches have been applied or proposed, like made-to-measure methods \cite{Syer:1996uv,Bovy:1704.03884}, a Jeans anisotropic modelling of disc stars \cite{Nitschai:1909.05269}, or analyses of halo stars \cite{Wegg_2019,Hattori:2012.03908}.
In the following subsections we discuss the main aspects of the most common methods.
We leave the discussion on the most interesting new ideas for section~\ref{sec:new-ideas}.

%%%%%%%%%%%%%%%%%%%%%%%%%%%%%%%%%%%%%%%%%%%%%%%%%%%%%%%%%%%%%%%%%%%%%%

\subsection{Modelling the distribution function}\label{sec:method:DF-modelling}

In spite of the hundreds of billions of stars that populate the Milky Way, their close encounters are so scarce that we can neglect any hard interaction between them, and treat the individual stars as tracer particles of a collisionless fluid, whose evolution is dictated by a continuous gravitational potential $\Phi$. For this reason, many studies of $\rhoDMlocal$ estimates revolve around solving the collisionless Boltzmann equation 
\begin{equation}\label{eq:Boltzmann-Cartesian}
\frac{\mathrm{d}f}{\mathrm{d}t} = \frac{\partial f}{\partial t} + \nabla_x f\cdot \mathbf{v} - \nabla_v f\cdot \nabla_x \Phi = 0
\end{equation}
for the distribution function $f(\mathbf{x},\mathbf{v},t)$ of a chosen tracer population, whose stars are located, at a given time $t$, at individual positions $\mathbf{x}$ with velocities $\mathbf{v}$.

Given the approximately axisymmetric shape of the luminous components of the Galaxy, it is convenient to express the Boltzmann equation \eqref{eq:Boltzmann-Cartesian} in Galactocentric cylindrical coordinates $\{R, \phi, z\}$ with associated velocities $\{ v_R, v_\phi, v_z \}$. Moreover, we use the following conventions: a positive $z$ points to the Galactic north and the Galaxy rotates clockwise. With this coordinate system, the Boltzmann equation is expressed as
\begin{equation}\label{eq:collisionless-Boltzmann-cylindrical}
\frac{\partial f}{\partial t}
+ p_R \frac{\partial f}{\partial R}
+ \frac{p_\phi}{R^2} \frac{\partial f}{\partial \phi}
+ p_z \frac{\partial f }{\partial z}
- \left( \frac{\partial \Phi}{\partial R} - \frac{p^2_\phi}{R^3} \right) \frac{\partial f}{\partial p_R}
- \frac{\partial \Phi}{\partial \phi} \frac{\partial f}{\partial p_\phi}
- \frac{\partial \Phi}{\partial z} \frac{\partial f}{\partial p_z}
= 0,
\end{equation}
where $p_R = v_R$, $p_\phi = R v_\phi$ and $p_z = v_z$.
Because of the time-derivative term, $\partial f / \partial t$, solving this equation is impractical, since any phase-space distribution of stars without observed accelerations could fit a given gravitational potential. 
Therefore, it is often necessary to assume that the Galaxy is dynamically old and therefore in a Galactic steady state, such that the $\partial f / \partial t$ term eq.~\eqref{eq:collisionless-Boltzmann-cylindrical} can be neglected altogether.

Any population of Galactic objects that is chosen as a tracer is finite in number and also associated with observational uncertainties.
If the full 6-dimensional phase space is to be fitted, observing a total of one million stars would only correspond to ten data points per dimension.
For this reason, in order to reduce the dimensionality of the problem so it is possible to solve eq.~\eqref{eq:collisionless-Boltzmann-cylindrical}, further considerations are taken.
Under the assumptions of the Ideal Galaxy, apart from the time-derivative term we can also drop the two terms that account for the azimuthal dependence, $(p_\phi/R^2)(\partial f/\partial \phi)$ and $(\partial \Phi/\partial \phi) (\partial f/\partial p_\phi)$, resulting in a simplified version of eq.~\eqref{eq:collisionless-Boltzmann-cylindrical}: 
\begin{equation}\label{eq:collisionless-Boltzmann-cylindrical-axisym-steady}
p_R \frac{\partial f}{\partial R}
+ p_z \frac{\partial f }{\partial z}
- \left( \frac{\partial \Phi}{\partial R} - \frac{p^2_\phi}{R^3} \right) \frac{\partial f}{\partial p_R}
- \frac{\partial \Phi}{\partial z} \frac{\partial f}{\partial p_z}
= 0.
\end{equation}
Another common approach is to assume a particular expression for $f(\mathbf{x},\mathbf{v},\bm{\vartheta})$ in terms of some parameters $\bm{\vartheta}$, flexible enough to encode all possible features of the real distribution, but constructed with a limited number of parameters that can be fitted to the available data.
The specific shape of $f$ depends on the tracer population it applies to (e.g., quasi-isothermal \cite{Dehnen:1999ea,Binney:2011xa} or exponential \cite{Vasiliev:1802.08239-agama} distribution functions for components of the disc, double-power-law \cite{Binney:2014jda,Posti:1411.7897} for spheroidal components, or an expression derived from a density profile---as the spherical case \cite{Cuddeford:1991MNRAS.253..414C}).

When defining a shape for $f$, a recently more popular practice is to express it in terms of the action-angle variables $\{\mathbf{J},\boldsymbol{\theta}\}$ \cite{BinneyTremaine:book,Binney:2013mhf,Sanders:1511.08213}, always assuming a Galactic steady state.
This choice is based on the Jeans theorem, that states that the distribution function of a stellar system where almost all orbits are regular can be expressed in terms of three independent integrals of motion, and those integrals can be the actions $\mathbf{J}$.
Hence, under appropriate assumptions, this leads to a feasible way of fitting the gravitational potential $\Phi$ and the distribution function $f$ of the tracer population. 

Once the gravitational potential has been inferred, the matter density distribution $\rho$ can be calculated via the Poisson equation, which in cylindrical coordinates takes the form
\begin{equation}\label{eq:Poisson-cylindrical}
\mathcal{R}
+ \frac{1}{R^2} \frac{\partial^2 \Phi}{\partial \phi^2}
+ \frac{\partial^2 \Phi}{\partial z^2}
= 4\pi G \rho,
\end{equation}
where
\begin{equation}\label{eq:rot-curve-term}
    \mathcal{R} = \frac{1}{R}\frac{\partial}{\partial R}\left( R\frac{\partial \Phi}{\partial R} \right)
\end{equation}
is commonly known as the `rotation curve' term, named after the definition of circular velocity (see eq.~\eqref{eq:Vc-definition}).
If the circular velocity is constant in $R$ around a position $R_0$, then $\mathcal{R}(R_0) = 0$ and the Poisson equation~\eqref{eq:Poisson-cylindrical} acquires a simpler form.
By combining eq.~\eqref{eq:Poisson-cylindrical} and the simultaneous fit of $\Phi$ and $f$ to the data, a value of $\rhoDMlocal$ can be derived from the fitted shape of the dark matter halo.

%%%%%%%%%%%%%%%%%%%%%%%%%%%%%%%%%%%%%%%%%%%%%%%%%%%%%%%%%%%%%%%%%%%%%%
\subsection{Moment methods}
A common way of estimating $\rhoDMlocal$ consists in modelling the first moments of the distribution function.
In this way one does not solve the Boltzmann equation~\eqref{eq:collisionless-Boltzmann-cylindrical} in its entirety, but rather its moment-based form leading to the Jeans equations.

In cylindrical coordinates, the three Jeans equations are obtained by multiplying eq.~\eqref{eq:collisionless-Boltzmann-cylindrical} times $v_R$, $v_z$ or $v_\phi$, followed by an integration over all velocities \cite{Jeans_1922,BinneyTremaine:book}. The respective Jeans equations are:
\begin{equation}\label{eq:1-R-Jeans-Cylindrical}
\frac{\partial \left( \nu \overline{v_R} \right)}{\partial t} 
+ \frac{\partial \left( \nu \overline{v^2_R} \right)}{\partial R}
+ \frac{\partial \left( \nu \overline{v_z v_R} \right)}{\partial z}
+ \frac{1}{R} \frac{\partial \left( \nu \overline{v_R v_\phi} \right)}{\partial \phi}
+ \nu \left( \frac{\overline{v^2_R} - \overline{v^2_\phi}}{R} + \frac{\partial \Phi}{\partial R} \right)
=0,
\end{equation}
\begin{equation}\label{eq:1-z-Jeans-Cylindrical}
\frac{\partial \left( \nu \overline{v_z} \right)}{\partial t} 
+ \frac{1}{R} \frac{\partial \left( R \nu \overline{v_z v_R} \right)}{\partial R}
+ \frac{\partial \left( \nu \overline{v^2_z} \right)}{\partial z}
+ \frac{1}{R} \frac{\partial \left( \nu \overline{v_z v_\phi} \right)}{\partial \phi}
+ \nu \frac{\partial \Phi}{\partial z}
=0,
\end{equation}
\begin{equation}\label{eq:1-phi-Jeans-Cylindrical}
\frac{\partial \left( \nu \overline{v_\phi} \right)}{\partial t} 
+ \frac{1}{R^2} \frac{\partial \left( R^2 \nu \overline{v_\phi v_R} \right)}{\partial R}
+ \frac{\partial \left( \nu \overline{v_\phi v_z} \right)}{\partial z}
+ \frac{1}{R} \frac{\partial \left( \nu \overline{v^2_\phi} \right)}{\partial \phi}
+ \frac{\nu}{R} \frac{\partial \Phi}{\partial \phi}
=0,
\end{equation}
where the number density is given by
\begin{equation}
\nu = \int \mathrm{d}^3 \mathbf{v} f(\mathbf{x},\mathbf{v},t)
\end{equation}
and an overlined quantity $\overline{A}$ represents the mean value of $A$ over velocity space:
\begin{equation}\label{eq:overline-A}
\overline{A} = \frac{1}{\nu}\int \mathrm{d}^3 \mathbf{v}\,A\, f(\mathbf{x},\mathbf{v},t).
\end{equation}
Notice that we are expressing the Jeans equations in terms of mean values, instead of moments of the distribution, but the two quantities are connected. The second moment $\sigma_{ij}$ of two velocity components $v_i$ and $v_j$, for example, corresponds to
\begin{equation}
    \sigma_{ij} = \frac{1}{\nu} \int \mathrm{d}^3 \mathbf{v} 
    \left( v_i - \bar{v}_i \right) \left( v_j - \bar{v}_j \right) f(\mathbf{x},\mathbf{v},t)
    = \overline{v_i v_j} - \bar{v}_i\, \bar{v}_j.
\end{equation}
Equations \eqref{eq:1-R-Jeans-Cylindrical}, \eqref{eq:1-z-Jeans-Cylindrical} and \eqref{eq:1-phi-Jeans-Cylindrical} are known as the radial, vertical and azimuthal Jeans equations, respectively.
Notice that the approximations of the Ideal Galaxy are not taken into account in the expressions that we present for these equations.

%%%%%%%%%%%%%%%%%%%%%%%%%%%%%%%%%%%%%%%%%%%%%%%%%%%%%%%%%%%%%%%%%%%%%%
\subsubsection{Vertical Jeans equation}\label{sec:method:vertical-Jeans}

An estimate of $\rhoDMlocal$ can be extracted from the vertical kinematics of disc stars, which oscillate in the vertical direction while they orbit the Galaxy. This oscillation is caused by the gravitational attraction of the mass---of both baryonic and dark matter origin---distributed close to the Galactic plane.
The local matter density can be inferred
either by solving the Boltzmann equation \eqref{eq:collisionless-Boltzmann-cylindrical} or from the vertical Jeans equation \eqref{eq:1-z-Jeans-Cylindrical}, which is the more common option. 
Afterwards, we can obtain $\rhoDMlocal$ after subtracting the contribution of baryons to the total local matter density.

Under the assumptions of the Ideal Galaxy, the vertical Jeans equation~\eqref{eq:1-z-Jeans-Cylindrical} is simplified to
\begin{equation}\label{eq:1-z-Jeans-Cylindrical-1D+tilt}
\mathcal{T}
+ \frac{\partial \left( \nu \overline{v^2_z} \right)}{\partial z}
+ \nu \frac{\partial \Phi}{\partial z}
=0,
\end{equation}
where
\begin{equation}\label{eq:tilt-term}
\mathcal{T} = \frac{1}{R} \frac{\partial \left( R \nu \overline{v_z v_R} \right)}{\partial R},
\end{equation}
commonly known as the `tilt' term, takes into account the correlation between the vertical and radial motions of the stars. 
This term, which is typically neglected, becomes increasingly important as we consider further distances from the Galactic plane. 
In ref.~\cite{Read:2014qva} it was estimated that $\mathcal{T}$ affects $\rhoDMlocal$ at $\sim 10\%$ for $|z| \lesssim 1\,\mathrm{kpc}$, indicating that $\mathcal{T}$ can be neglected close to the Galactic plane.

The vertical extent of the data is a key aspect in this type of analysis.
On the one hand, staying at a closer distance to the Galactic plane allows us to neglect $\mathcal{T}$ and to assume a constant dark matter density.
On the other hand, this type of $\rhoDMlocal$ studies tend to perform better if the data reaches further from the Galactic plane, because dark matter dominates the matter density at greater heights, making it less degenerate with the baryonic matter distribution.
However, both the physical extension of the disc and, more importantly, observational limitations at further distances, makes it very difficult to go beyond $|z| \sim 1.5\,\mathrm{kpc}$. A notable exception is the recent study of ref.~\cite{Salomon:2020eer}.

Another aspect to consider about the vertical range of the data is the relaxation time, which is important for the validity of the steady-state assumption.
Stars with lower vertical energy have a higher vertical oscillation frequency, and can be expected to recover a steady-state configuration faster after being perturbed.

All in all, $\rhoDMlocal$ can be estimated by solving the vertical Jeans equation \eqref{eq:1-z-Jeans-Cylindrical-1D+tilt} together with the Poisson equation \eqref{eq:Poisson-cylindrical} for a specific model of the local mass distribution.
This type of analysis can be simplified to only one dimension provided that we can a) assume a constant dark matter density; b) neglect the tilt term; and c) consider a flat circular velocity in the Sun's vicinity, around a Galactocentric radius $R_\odot$.
The first two conditions reduce the vertical Jeans equation~\eqref{eq:1-z-Jeans-Cylindrical-1D+tilt} to one dimension, and the third condition does the same for the Poisson equation~\eqref{eq:Poisson-cylindrical}.

%%%%%%%%%%%%%%%%%%%%%%%%%%%%%%%%%%%%%%%%%%%%%%%%%%%%%%%%%%%%%%%%%%%%%%
\subsubsection{Fitting the circular velocity curve}\label{sec:method:Vc-fitting}
An estimate of $\rhoDMlocal$ can also be obtained from the circular velocity curve of the Milky Way.
Stars that move through the Galactic plane in close-to-circular orbits have an azimuthal velocity that depends on the total matter enclosed within the Galactic radius $R$ of their orbit. 
Hence, from the observation of the circular velocity $v_{\rm c}(R)$ at different radii, it is possible to extract the matter profile of the Galaxy. 
In fact, one of the first indications for the need of dark matter came from the study of galactic rotation curves, as e.g. in the works of Knut Lundmark \cite{Lundmark:1930} and Vera Rubin \cite{Rubin:1980zd}.

From the gravitational potential $\Phi$ set by a given Galactic model, the circular velocity at a certain radius can be extracted through the relation
\begin{equation}\label{eq:Vc-definition}
v^2_{\rm c}(R) = R \left.\frac{\partial \Phi}{\partial R}\right|_{z=0}.
\end{equation}
Thus, the parameters of a given potential $\Phi$ can be fitted from a comparison of the theoretical $v_{\rm c}(R)$ from eq.~\eqref{eq:Vc-definition} to its observed values $v_{\rm c,obs}(R)$, which can be computed, e.g., from the radial Jeans equation \eqref{eq:1-R-Jeans-Cylindrical}. 
A recent review \cite{Sofue:2020rnl} discusses different approaches to obtain Galactic circular velocity measurements, including a list of $\rhoDMlocal$ estimates from fits to the circular velocity curve of the Milky Way.

Under the considerations of the Ideal Galaxy---in particular assuming a steady state and axisymmetry---and neglecting the term $\partial \left( \nu \overline{v_z v_R} \right)/\partial z$ in eq.~\eqref{eq:1-R-Jeans-Cylindrical}, which only has an $\mathcal{O}(1\%)$ effect on $v_{\rm c}$ at radii up to $\sim 20 \,\mathrm{kpc}$ \cite{Eilers:1810.09466}, the radial Jeans equation \eqref{eq:1-R-Jeans-Cylindrical} can be rewritten as
\begin{equation}\label{eq:1-R-Jeans-Cylindrical-simplified}
v_{\rm c,obs}^2 = 
\overline{v^2_\phi}
- \overline{v^2_R} \left(
1 + \frac{\partial\ln\nu}{\partial\ln R} + \frac{\partial\ln \overline{v^2_R}}{\partial\ln R}
\right).
\end{equation}

One thing that becomes clear from eq.~\eqref{eq:1-R-Jeans-Cylindrical-simplified} is the fact that $v_{\rm c}$ and $v_\phi$ are not necessarily the same.
This reflects what is called the `asymmetric drift' \cite{BinneyMerrifield:book}, an observable indication that stellar orbits are not exactly circular.
More precisely, the asymmetric drift implies that the average rotational velocity $\sqrt{\overline{v^2_\phi}}$ is typically smaller than the circular velocity $v_{\rm c}$, which is partly explained because Galactic matter densities grow as we move towards their centres, meaning that there are more stars whose azimuth velocity $v_\phi < v_{\rm c}$ than the amount of stars with $v_\phi > v_{\rm c}$ \cite{BinneyTremaine:book}.

In the same way as when solving the vertical Jeans equation~\eqref{eq:1-z-Jeans-Cylindrical-1D+tilt}, the radial Jeans equation~\eqref{eq:1-R-Jeans-Cylindrical-simplified} provides a way of constraining the gravitational potential $\Phi$. 
From this point, one can use Poisson equation \eqref{eq:Poisson-cylindrical} to infer $\rhoDMlocal$ for a suitable Galactic mass model. 

%%%%%%%%%%%%%%%%%%%%%%%%%%%%%%%%%%%%%%%%%%%%%%%%%%%%%%%%%%%%%%%%%%%%%%

\subsection{Spatial coverage}\label{sec:spatial-coverage}

A key aspect that distinguishes between different approaches to get $\rhoDMlocal$ is the volume sampled by the data. Depending on the final goal of the study, one might prefer to use data widely spread in the Galaxy, located in the Galactic halo, distributed in the Galactic plane or concentrated to the local neighbourhood of the Sun.
Studies covering different volumes are complementary, in the sense that they suffer from different biases, and can make use of different simplifying assumptions. 
We can differentiate between \emph{global} and \emph{local} analyses, separating the $\rhoDMlocal$ studies in a similar way to ref. \cite{Read:2014qva}. 

\medskip

\noindent\textbf{Global analyses.} They offer an average value of $\rhoDMlocal$ over a large volume, making them particularly useful, for example, for some indirect dark matter searches, where global properties of the dark matter halo are typically needed instead of the specific values at our local environment. 

\medskip

\noindent\textbf{Local analyses.} They aim at obtaining $\rhoDMlocal$ at 
a comparatively small volume around the Sun's location and are typically less dependent on assumptions pertaining to larger spatial scales.
This quantity is important, for example, in dark matter direct detection experiments, where the truly local density of dark matter is what would enable an interpretation of a positive signal. 

\medskip

\noindent Although it seems easy to categorise an analysis either as global or local, the distinction between global and local analyses is not always clear. For example, studies using data from a local spatial volume might be complemented by constraints and assumptions coming from observations on a larger spatial scale.
We prefer to adopt the term \emph{global} for such studies, based on the Galactic volume that is ultimately included in the analyses, but they could also fit in the local category judging by their main tracers' location. 

An advantage of global methods with respect to local studies is that, since their tracers cover a larger volume, spatial fluctuations in their properties---such as their number density---are averaged out. However, the downside of this averaging---in terms of finding $\rhoDMlocal$---is that it misses possible interesting local features, such as signatures of disequilibria.
The advantage of local studies over global studies is their enhanced sensitivity to local density fluctuations, which makes their resulting $\rhoDMlocal$ closer to the actual value at the Sun's Galactic position. Typical problems associated to local studies come from the lack of statistics or the non-validity of one of the assumptions.
Recent surveys have significantly improved on the amount of data available over the last few years. 
In particular, the Gaia satellite has produced astrometric measurements for more than a billion objects, with a parallax precision as low as $\sim 10 \,\mathrm{\mu as}$ for the best objects in Gaia's early third data release (EDR3) \cite{Brown:2012.01533}.
Its completeness is strongly dependent on stellar crowding, but for areas of the sky with source densities smaller than $2\times 10^5~\mathrm{deg}^{-2}$, this data set is practically complete for apparent magnitudes $G \lesssim 20$ \cite{Fabricius:2012.06242}.
However, these improvements do not, in and of themselves, address the problem of common assumptions and their validity, such as a locally flat rotation curve or a mirror symmetry across the Galactic plane. On the contrary, a shrinking statistical uncertainty demands a more careful and accurate treatment of systematic uncertainties. 

Differences between the estimated value of $\rhoDMlocal$ from global and local analyses can give us further information about our Galaxy. For example, a larger value from a local study could be an indication of an oblate Galactic halo if the global study assumed the halo to be spherical \cite{Read:2014qva}. 
We must note, however, that current deviations in the results between local and global methods are typically well contained within the uncertainties of the estimates, and any possible discrepancy should disappear within the error range if systematic uncertainties, due to e.g. mismodelling, are also taken into account. After all, if one had a perfect model of the Galaxy, the same conclusion should be reached regardless of the method, and the same value of $\rhoDMlocal$ ought to be recovered within the uncertainty of each study. 

%%%%%%%%%%%%%%%%%%%%%%%%%%%%%%%%%%%%%%%%%%%%%%%%%%%%%%%%%%%%%%%%%%%%%%
%%%%%%%%%%%%%%%%%%%%%%%%%%%%%%%%%%%%%%%%%%%%%%%%%%%%%%%%%%%%%%%%%%%%%%
\section{Estimates of $\rhoDMlocal$}\label{sec:rhoDM-estimates}

The very first dynamical mass measurements of the solar neighbourhood were made roughly a century ago, by Kapteyn \cite{Kapteyn:1922zz}, Jeans \cite{Jeans_1922}, and Oort \cite{Oort:1932_a}, using the vertical motion of stars. By comparing their rough estimates with the amount of visible stars, they placed constraints on the local density of non-luminous matter (e.g., gas, dust and planets).
Significant progress was made in the late 1980's by Kuijken and Gilmore \cite{Kuijken:1989_a,Kuijken:1989hu,Kuijken:1989_c,Kuijken:1991}, who measured the total matter density of the solar neighbourhood to roughly $0.1\,\mathrm{M_\odot / pc^3}$, consistent with the observed densities of stars and gas.
In recent decades, local mass measurements have became all the more refined with further improvements to the quality and quantity of data, not least in the late 1990's with ESA's astrometric satellite mission Hipparcos \cite{Perryman:1997sa}, and estimates of $\rhoDMlocal$ converged to a value close to $\rhoDMlocal \sim 0.4\,\mathrm{GeV/cm^3}$. An instructive review on such estimates up until 2013 is presented in ref.~\cite{Read:2014qva}. Our goal is not to repeat the discussion of that review, but rather to focus on studies published after 2013.

Many studies have been published in the last years, especially after Gaia's second data release (DR2) \cite{Brown:2018dum}.
In the next subsections we describe the progress made by recent studies.
We start in section~\ref{sec:rhoDM:local-and-very-local} with the works that only include information from local and very local observations, which do not fit a global Galactic mass model.
In section~\ref{sec:rhoDM:rot-curve} we discuss those works that fitted the circular velocity curve of the Milky Way, but included little or no additional Galactic information in the analyses.
In section~\ref{sec:rhoDM:DF-global} we discuss the progress made by other global analyses, which typically combine several observations, including a few studies based on either disc stars, halo stars, or circular velocity data.
Finally, in section~\ref{sec:rhoDM:summary} we summarise the results and conclusions of all studies included in the previous subsections. 
We leave the discussion of future estimates for section~\ref{sec:new-ideas}.

%%%%%%%%%%%%%%%%%%%%%%%%%%%%%%%%%%%%%%%%%%%%%%%%%%%%%%%%%%%%%%%%%%%%%%
\subsection{From local and very local observations}\label{sec:rhoDM:local-and-very-local}

Studying $\rhoDMlocal$ from local observations is perhaps the most common
way of estimating the value of this parameter. 
Local studies most often employ the vertical Jeans equation (see section \ref{sec:method:vertical-Jeans}), and sometimes the method of distribution function fitting (see section \ref{sec:method:DF-modelling}).
The works discussed in this section do not rely on prior information from a large volume outside a few kiloparsecs of the Sun's location, thus
benefiting from the advantages of a local study (see section \ref{sec:spatial-coverage}). 
One such benefit is that the modelling is sensitive to the dark matter density specifically at the Sun's location, unlike global studies. 

%%%%%%%%%%%%%%%%%%%%%%%%%%%%%%%%%%%%%%%%%%%%%%%%%%%%%%%%
%%%
%%%           TABLE with LOCAL (beginning)
%%%
%%%%%%%%%%%%%%%%%%%%%%%%%%%%%%%%%%%%%%%%%%%%%%%%%%%%%%%%

\begin{figure}
\noindent{\footnotesize
\begin{minipage}[t]{\textwidth}
\renewcommand\footnoterule{}
\captionof{table}{Recent $\rhoDMlocal$ estimates from studies based on the kinematics of local and very local observations discussed in section~\ref{sec:rhoDM:local-and-very-local}. 
Distances are expressed in units of $\mathrm{kpc}$ and $\rhoDMlocal$ in units of $\mathrm{GeV/cm^{3}}$. $R'$ is the Heliocentric cylindrical radius.
Errors are $1\sigma$ unless stated otherwise.
Labels of the local analyses correspond to the values represented in figure~\ref{fig:rhoDM-all-vertical} in brown colour and with triangular markers.
\label{Tab:rhoDM-values-local}}
\centering
\begin{tabular}{lcllcr}
\hline\\[-2ex]
\multicolumn{4}{l}{\qquad Local vertical kinematics analyses}\\
\hline\\[-2ex]
\textbf{Label} & $\boldsymbol z$ \textbf{range} & \textbf{Tracer description} & $\boldsymbol \rhoDMlocal$ & \textbf{Ref.}\\
\hline\\[-2ex]
% McKee+15
McKee+15
& --- 
& comprehensive list
& $0.49\pm 0.13$ 
& \cite{McKee:2015hwa}
\\
& & of star and gas \\
& & observations \\
\hline\\[-2ex]
% Xia+16
Xia+16
& $0.2 < z < 1.5$
& 1427 G \& K stars
& $0.60^{+0.18}_{-0.21}$ 
& \cite{Xia:2015agz}
\\
& & (LAMOST DR2)\\
\hline\\[-2ex]
% Sivertsson+18
Sivertsson+18\footnote{The quoted $\rhoDMlocal$ values correspond to the analyses of their $\alpha$-young (y) and $\alpha$-old (o) populations, with and without the inclusion of the tilt term. 
The authors consider the $\alpha$-young population, with tilt term included, to give the most reliable result; this is based on large discrepancies in the $\alpha$-old analyses, that could be due to systematic uncertainties associated with a mismodelling of the tilt term or disequilibrium effects.
}
& $|z| \sim \text{0.5--1.2}$ (y) 
& G dwarfs from \cite{Budenbender:2014xra}
& $0.46^{+0.07}_{-0.09}$ (y - tilt) & \cite{Sivertsson:2017rkp}
\\
& $|z| \sim \text{0.6--2.3}$ (o) & (SDSS/SEGUE) & $0.48^{+0.05}_{-0.06}$ (y - no tilt) \\
& & & $0.73^{+0.06}_{-0.05}$ (o - tilt) \\
& & & $0.46^{+0.02}_{-0.02}$ (o - no tilt) \\
& & & $0.40^{+0.03}_{-0.03}$ (y+o - tilt) \\
\hline\\[-2ex]
% Hagen+18
Hagen+18
& $|z| \sim \text{0.6--1.5}$
& red clump stars
& $0.68\pm 0.08$ & \cite{Hagen:1802.09291}
\\
& & ($\text{TGAS}\times \text{RAVE DR5}$)\\
\hline\\[-2ex]
% Guo+20
Guo+20\footnote{The quoted $\rhoDMlocal$ values correspond to the analyses including data from the Galactic north (N) and south (S), with a Gaussian prior on either the total stellar surface density, $P(\Sigma_*)$, or the stellar volume density at $z=0$, $P(\rho_{*,0})$. 
The authors consider the results from their N+S analysis with $P(\Sigma_*)$ to be the most reliable.
}
& $0 < |z| < 1.3$
& $\sim 90\,000$ G \& K stars
& $0.50_{-0.08}^{+0.09}$ (N+S, $P(\Sigma_*)$) & \cite{Guo:2020rcv}
\\
& & ($\text{LAMOST DR5}\times \text{Gaia DR2}$) & $0.27_{-0.16}^{+0.22}$ (N+S, $P(\rho_{*,0})$) \\
& & & $0.65_{-0.08}^{+0.08}$ (N, $P(\Sigma_{*})$) \\
& & & $0.19_{-0.12}^{+0.15}$ (S, $P(\Sigma_{*})$) \\
\hline\\[-2ex]
% Salomon+20
Salomon+20\footnote{The quoted $\rhoDMlocal$ values correspond to the Galactic north (N) and south (S) analyses.}
& $0.6 < |z| < 3.5$
& $43\,589$ red clump stars
& $0.51 \pm 0.09$ (N) & \cite{Salomon:2020eer}
\\
& & ($\text{Gaia DR2}$) & $0.37 \pm 0.09$ (S) \\
\hline\\[-2ex]
\multicolumn{4}{l}{\qquad Very local analyses}\\
\hline\\[-2ex]
\textbf{Label} & \textbf{Spatial cuts} & \textbf{Tracer description} & $\boldsymbol \rhoDMlocal$ & \textbf{Ref.}\\
\hline\\[-2ex]
% Schutz+18
Schutz+18
& $|z|<0.2$
& 1599 A stars
& $1.43^{+0.45}_{-0.56}$ (A stars)
& \cite{Schutz:2017tfp}
\\
& $\Delta R' < 0.15$ 
& $16\,302$ F stars
& $0.71^{+0.45}_{-0.41}$ (F stars)
\\
&
& $14\,252$ G stars
& $0.15^{+0.38}_{-0.15}$ (G stars)
\\
& & (Gaia DR1 [TGAS])\\
\hline\\[-2ex]
% Buch+19
Buch+19
& $|z|<0.2$
& 4445 A stars
& $0.61^{+0.38}_{-0.38}$ (A stars) 
& \cite{Buch:2018qdr}
\\
& $\Delta R' < 0.15$ 
& $37\,707$ F stars
& $1.48^{+0.30}_{-0.30}$ (F stars)
\\
&
& $43\,332$ G stars
& $0.42^{+0.38}_{-0.34}$ (G stars)
\\
& & (Gaia DR2)\\
\hline
\hline
\end{tabular}
\vspace{-2ex}
\end{minipage}
}
\end{figure}

%%%%%%%%%%%%%%%%%%%%%%%%%%%%%%%%%%%%%%%%%%%%%%%%%%%%%%%%
%%%
%%%           TABLE with LOCAL (end)
%%%
%%%%%%%%%%%%%%%%%%%%%%%%%%%%%%%%%%%%%%%%%%%%%%%%%%%%%%%%

Table~\ref{Tab:rhoDM-values-local} shows the $\rhoDMlocal$ estimates from local and very local studies. In general, the result of the studies that are in the same category agree well with each other, with only a few exceptions.
In the first part of the table, we include as \emph{local} studies those that are based on observations of disc stars that extend up to $|z| \simeq 1.5\,\mathrm{kpc}$, as well as ref.~\cite{Salomon:2020eer}, which goes up to $|z| \simeq 3.5\,\mathrm{kpc}$. 
Close to the Galactic plane and stellar disc, the total matter density is dominated by baryons; therefore, in order to break the degeneracy between baryons and dark matter, it is useful to reach greater heights---the maximum height of the considered spatial volume has mostly been limited by availability and quality of the data itself.
However, a couple of studies have attempted to estimate $\rhoDMlocal$ from observations that are enclosed within $\mathcal{O}(100\,\mathrm{pc})$ from the Sun's location.
These studies, which we categorise as \emph{very local}, are included in the second part of table~\ref{Tab:rhoDM-values-local}.
There are a few reasons why it is interesting to test the matter density profile so close to the location of the Sun.
On the one hand, the very local mass distribution could differ significantly from the surrounding distribution as a consequence of non-equilibrium effects (e.g., \cite{Widmark:2020vqi}). 
On the other hand, such a local study enables us to test several non-standard hypothesis, like the possible existence of a thin dark matter disc.
An extended discussion on both scenarios is provided in section~\ref{sec:breaking-of-Ideal-Galaxy}.
The drawback of staying this close to the Sun is that it exacerbates the difficulty of isolating the dark matter contribution to the gravitational potential, given the even larger fraction of baryons.
This obstacle explains the larger uncertainties in the very local $\rhoDMlocal$ inferences with respect to the local inferences in the studies included in table~\ref{Tab:rhoDM-values-local}. 

\subsubsection{Studies based on local observations}\label{sec:rhoDM:local-and-very-local:local}

We start by describing the works included in the first part of table~\ref{Tab:rhoDM-values-local}, which extend further in the vertical direction while still remaining local. All studies presented in this section used the vertical Jeans equation in their main analyses, but they followed different strategies to infer $\rhoDMlocal$. The incompatibility of some results could be caused by the consideration of different assumptions.
For example, ref.~\cite{Sivertsson:2017rkp} considered two independent tracer populations, finding $\rhoDMlocal$ values that can be very different depending on whether the tilt term is included in their Jeans analysis.
Another example is the disagreement in the resulting $\rhoDMlocal$---found in refs.~\cite{Guo:2020rcv,Salomon:2020eer}---when independently analysing the north and south Galactic hemispheres.
These differences could contain useful information about the current configuration of the Milky Way, and are a clear indication that we need to move beyond many of the common assumptions and simplifications, like that of a steady state or neglecting the tilt term, in order to increase both the precision and the accuracy of $\rhoDMlocal$ estimates. In section \ref{sec:breaking-of-Ideal-Galaxy} we present a discussion on this subject. 

In the following paragraphs we are going to describe the main aspects of the works presented in the local category of table \ref{Tab:rhoDM-values-local}.

Starting with ref.~\cite{McKee:2015hwa}, the authors analysed the surface density and vertical distribution of baryons at the Sun's position from a comprehensive list of sources---including several types of stars as well as molecular, atomic and ionised gas---from which they estimated the baryonic surface density $\Sigma_{\rm b}$ at $|z| = 1.0 \,\mathrm{kpc}$ and $|z| = 1.1 \,\mathrm{kpc}$.
Comparing their value of $\Sigma_{\rm b}$ with the total surface density from the works of \cite{Kuijken:1991,Bovy:2013raa,Zhang:2012rsb,Bienayme:2014kva}, they derived a value of $\rhoDMlocal = (0.49\pm 0.13)\,\mathrm{GeV/cm^3}$.

Reference \cite{Xia:2015agz} estimated $\rhoDMlocal$ from G and K type stars located in the Galactic north, based on LAMOST DR2 observations, at a vertical distance from the Galactic plane between $0.2\,\mathrm{kpc}$ and $1.5\,\mathrm{kpc}$.
The spatial volume of their data is in the shape of a cone centred on the Sun's position, defined by a latitude angle of $b > 85^\circ$.
They obtained $\rhoDMlocal$ by fitting a local vertical mass model that assumes a constant dark matter density, an exponential stellar density, and a razor thin gas contribution.
Neglecting the tilt term and assuming a flat rotation curve, they quoted a value of $\rhoDMlocal = 0.60_{-0.21}^{+0.18}\,\mathrm{GeV/cm^3}$.

The authors of \cite{Sivertsson:2017rkp} applied a method presented in \cite{Silverwood:2015hxa} to G dwarf stars processed in \cite{Budenbender:2014xra}, divided in two tracer populations: an $\alpha$-young set of stars defined to have metallicity cuts $\mathrm{[\alpha/Fe]} < 0.2$ and $-0.5 < \mathrm{[Fe/H]}$, and an $\alpha$-old population with metallicity cuts $0.3 < \mathrm{[\alpha/Fe]}$ and $-1.2 < \mathrm{[Fe/H]} < -0.3$.
The vertical extension of the $\alpha$-young and $\alpha$-old populations is $|z| \sim \text{0.5--1.2}\,\mathrm{kpc}$ and $|z| \sim \text{0.6--2.3}\,\mathrm{kpc}$, respectively.
The method used in this paper is based on fitting the first moments of the tracer's distribution function to the data. They assumed a specific vertical mass model, which consists on a constant dark matter density and a baryonic distribution based on refs.~\cite{McKee:2015hwa,Zheng:2001wc,Flynn:2006tm}.
They present the results of five analyses in which axisymmetry is assumed: two analyses of the $\alpha$-young population, one of which includes the contribution from the tilt term while the other analysis does not; two analogue analyses of the $\alpha$-old population, and a fifth analysis combining the $\alpha$-young and $\alpha$-old populations accounting for the tilt term.
The results coming from the two populations differ when the tilt term is included---see the values listed in table \ref{Tab:rhoDM-values-local}. 
The authors quoted $\rhoDMlocal = 0.46_{-0.09}^{+0.07}\,\mathrm{GeV/cm^3}$ as their preferred estimate, which results from their $\alpha$-young analysis including the tilt term. 

In ref.~\cite{Hagen:1802.09291}, the authors derived $\rhoDMlocal$ using red clump stars from TGAS (Tycho-Gaia Astrometric Solution) cross-matched with RAVE DR5 observations. The volume studied to obtain $\rhoDMlocal$ covers $|z| \sim \text{0.6--1.5}\,\mathrm{kpc}$ and a Galactocentric radius of $|R- R_\odot| < 0.5 \,\mathrm{kpc}$.
Similarly to ref.~\cite{Sivertsson:2017rkp}, they analysed two independent samples divided by metallicity: a metal rich sample associated with the thin disc and a metal poor sample associated with the thick disc. In their statistical analysis, the observed surface density in the studied $z$ range is compared to the theoretical values from their mass model.
Assuming double exponential discs for the thin and thick stellar components, a razor thin gas disc and a constant dark matter contribution, they obtained a value of $\rhoDMlocal = (0.68\pm 0.08)\,\mathrm{GeV/cm^3}$. However, the quoted uncertainty does not take into account the impact of varying the scale heights of the stellar discs, which they show that can easily affect $\rhoDMlocal$ at a 30\% level with only a 10\% change in the thin disc scale height.

The study presented in \cite{Guo:2020rcv} estimated $\rhoDMlocal$ from an analysis that closely follows that of \cite{Xia:2015agz}, with different spatial cuts and larger statistics.
The studied data sample is derived from a cross-match between Gaia DR2 and LAMOST DR5 observations, amounting to a total of $93\,609$ G and K type dwarf stars, almost 2 orders of magnitude larger than the data sample used in ref.~\cite{Xia:2015agz}.
The selection cuts applied for their main analysis include the spatial region, in Galactocentric cylindrical coordinates, defined by $0 < |z| < 1.3 \,\mathrm{kpc}$, $|R- R_\odot| < 0.2 \,\mathrm{kpc}$ and $|\phi| < 5^\circ$, with the additional constraint over the distance from the Sun's location of $d_\odot > 0.2 \,\mathrm{kpc}$ to avoid selection effects in the bright end.
Apart from the larger data set, the main difference in this work with respect to ref.~\cite{Xia:2015agz} is the consideration of a Gaussian prior for the total stellar surface density in their main analysis, in which they neglected the tilt and rotation curve terms, and assumed a constant dark matter density, a razor thin gas disc and a single stellar (thin) disc. 
From this configuration they quoted a value of $\rhoDMlocal = 0.50_{-0.08}^{+0.09}\,\mathrm{GeV/cm^3}$, but they studied the dependence of the inferred $\rhoDMlocal$ on different prior and selection-cut choices, whose main conclusions will be addressed in section~\ref{sec:diff-between-estimates}.
Relevant examples are included in table~\ref{Tab:rhoDM-values-local}.

The authors of ref.~\cite{Salomon:2020eer} developed an analysis that includes observations of red clump stars from Gaia DR2. The studied data reaches a larger $|z|$ than in the other works shown in table~\ref{Tab:rhoDM-values-local}, covering a vertical range between $0.6\,\mathrm{kpc}$ and $3.5\,\mathrm{kpc}$ both to the Galactic north and south.
The analysed regions within each Galactic hemisphere are further divided into five sub-regions parallel to the Galactic plane: one including the location of the Sun, two in the directions of the Galactic centre and anti-centre, and two in the azimuthal directions.
A detailed picture of the sub-regions extension can be seen in figure~3 of \cite{Salomon:2020eer}.
Since their data extends quite far from the Galactic plane, they accounted for the tilt term, as well as the rotation term correction as a function of $R$ and $z$.
They found that the Galactic north is more disturbed than the south, which is reflected in the $\rhoDMlocal$ estimates.
They inferred a value of $\rhoDMlocal = (0.5087 \pm 0.0909) \mathrm{GeV/cm^3}$ for the north sample and $\rhoDMlocal = (0.3736 \pm 0.0871) \mathrm{GeV/cm^3}$ for the south sample.
In order to obtain $\rhoDMlocal$, they assumed that the surface density at $|z| = \text{2--3.5}\,\mathrm{kpc}$ is entirely dominated by dark matter.
As an additional test, they performed a distribution function fitting analysis, closely following the method of \cite{Bienayme:2014kva}.
For this analysis they did not separate the north and south observations, and they obtained a range of $\rhoDMlocal \sim \text{0.41--0.54}\,\mathrm{GeV/cm^3}$, largely dependent on the value of $v_{\rm c} (R_\odot)$ and its radial derivative. Although this range is compatible with the inferred values from their Jeans analyses, they caution the reader about the compatibility of both results, since they derive a larger $\rhoDMlocal$ from the distribution function fitting technique when they apply the same $v_{\rm c} (R)$ constraints as in the Jeans analyses.

\subsubsection{Studies based on very local observations}\label{sec:rhoDM:local-and-very-local:very-local}

In this section we summarise the very local analyses included in table~\ref{Tab:rhoDM-values-local}, which covered spatial volumes whose vertical extent is only a few hundred parsecs, typically $|z| < 200\,\mathrm{pc}$. The main purpose of most of these papers was to constrain the possibility of a thin dark disk, which is discussed in more detail in section~\ref{sec:breaking-of-Ideal-Galaxy}. However, in the absence of a dark disk, refs. \cite{Schutz:2017tfp,Buch:2018qdr} also included estimates of $\rhoDMlocal$.

In volumes this close to the Galactic plane, baryonic matter dominates, while dark matter probably constitutes no more than $\sim 10\%$ of the total mid-plane matter density. Therefore, estimating $\rhoDMlocal$ necessitates a precise understanding of the baryonic components, as well as a precise measurement of the total matter density.

Most papers that analysed the very local volume used the same baryonic model \cite{McKee:2015hwa}, based on different pre-Gaia studies \cite{Flynn:2006tm,Kramer:2016dew}. Important systematic uncertainties can be associated with this choice, since this model presents some issues when predicting the vertical matter density. A thorough discussion on the baryonic model is presented in section \ref{sec:importance-baryons}.

For measurements of the total matter density, precision has increased significantly during the Gaia era.
Because of the lack of radial velocities, earlier Gaia-based studies were limited to using the velocity information of stars with small Galactic latitudes, for which the vertical velocity was given by parallax and proper motion \cite{Widmark:1711.07504,Schutz:2017tfp,Buch:2018qdr}. The availability of radial velocities with Gaia DR2 made it possible to use the velocity information of almost all stars in the studied volume, which has granted significant power of inference \cite{Widmark:2018ylf,Widmark:2020vqi}.

References \cite{Schutz:2017tfp} and \cite{Buch:2018qdr} are similar, in terms of their modelling, choice of stellar tracer populations (A, F, and G type stars), and spatial volume (a cylinder defined by $\Delta R' < 150\,\mathrm{pc}$ and $|z| < 200\,\mathrm{pc}$, where $R'$ is the solar-centered radius). 
In ref. \cite{Schutz:2017tfp}, the authors obtained the following result for their three stellar samples, using Gaia DR1: $\rhoDMlocal = 1.43^{+0.45}_{-0.56}\,\mathrm{GeV/cm^3}$ for A type stars; $\rhoDMlocal = 0.71^{+0.45}_{-0.41} \,\mathrm{GeV/cm^3}$ for F type stars; and $\rhoDMlocal = 0.15^{+0.38}_{-0.15}\,\mathrm{GeV/cm^3}$ for early G type stars.
The large uncertainties in their $\rhoDMlocal$ estimates are not surprising, given the close distance to the Galactic plane and the limited Gaia DR1 data set. What is impressive is the fact that they were able to obtain credibility intervals for $\rhoDMlocal$ that did not entirely cover their chosen prior range.
In ref. \cite{Buch:2018qdr}, the authors used Gaia DR2, which granted them an improved volume completion of $\sim 80\%$, compared to $\sim 30\%$ from Gaia DR1 given their selection cuts. 
They obtained the following results: $\rhoDMlocal = 0.61^{+0.38}_{-0.38}\,\mathrm{GeV/cm^3}$ for the A type stars; $\rhoDMlocal = 1.48^{+0.30}_{-0.30}\,\mathrm{GeV/cm^3}$ for the F type stars; and $\rhoDMlocal = 0.42^{+0.38}_{-0.34}\,\mathrm{GeV/cm^3}$ for the early G stars. The credibility intervals are smaller than for ref. \cite{Schutz:2017tfp}; still, there are significant discrepancies between the results of the different stellar tracer populations, which is unlikely to be explained solely by statistical variance.

A list of studies similarly close to the location of the Sun were carried out in \cite{Widmark:1711.07504} using Gaia DR1, and in \cite{Widmark:2018ylf,Widmark:2020vqi} using Gaia DR2. The dynamical models of these studies are independent of the baryonic model, such that the gravitational potential and matter density is quite free to vary in both amplitude and shape. The inferred results of refs. \cite{Widmark:2018ylf,Widmark:2020vqi} were compared to the baryonic model post-inference, and both found a large excess of mass very close to the Galactic mid-plane. However, the latter study \cite{Widmark:2020vqi}, which went further in height ($|z| < 400\,\mathrm{pc}$), showed that these results are likely due, at least in part, to time-varying dynamical effects. Because of the spurious results of these studies, which are indicative of large systematic biases, no value for $\rhoDMlocal$ was presented. Further discussion on these papers will be given in section~\ref{sec:breaking-of-Ideal-Galaxy}.

%%%%%%%%%%%%%%%%%%%%%%%%%%%%%%%%%%%%%%%%%%%%%%%%%%%%%%%%%%%%%%%%%%%%%%
\subsection{From the Galactic circular velocity curve}\label{sec:rhoDM:rot-curve}

An estimate of $\rhoDMlocal$ can be obtained by fitting the Milky Way's circular velocity curve to a specific Galactic mass model (see section~\ref{sec:method:Vc-fitting}).
This method, applied to external galaxies, provided one of the first indications for the existence of dark matter \cite{Rubin:1980zd}. However, given our specific location inside the Milky Way, it is difficult to obtain precise observations covering a wide range of distances from the Galactic centre. In particular, the presence of intergalactic dust and gas can make it more difficult to observe objects within our own Galactic disc than to measure the rotation curve of some external galaxies.
In spite of this handicap, astronomers have managed to measure the Galactic circular velocity curve up to Galactocentric distances of $\sim 100 \,\mathrm{kpc}$ \cite{Sofue:2020rnl}. 
In addition, circular velocities in the local region, covering distances within $4\,\mathrm{kpc}\lesssim R \lesssim 25\,\mathrm{kpc}$, have been especially well determined thanks to the observations available after Gaia DR2 \cite{Eilers:1810.09466,Mroz:1810.02131}.

%%%%%%%%%%%%%%%%%%%%%%%%%%%%%%%%%%%%%%%%%%%%%%%%%%%%%%%%
%%%
%%%           TABLE with ROT. CURVE (beginning)
%%%
%%%%%%%%%%%%%%%%%%%%%%%%%%%%%%%%%%%%%%%%%%%%%%%%%%%%%%%%

\begin{figure}
\noindent{\footnotesize
\begin{minipage}{\textwidth}   % GUESS (OR CALCULATE) MINIPAGE WIDTH
\renewcommand\footnoterule{}     % ELIMINATE LITTLE LINE SEPARATER
\captionof{table}{Recent $\rhoDMlocal$ estimates from studies that fitted the circular velocity curve of the Galaxy, described in section~\ref{sec:rhoDM:rot-curve}.
Distances are given in units of $\mathrm{kpc}$ and $\rhoDMlocal$ in units of $\mathrm{GeV/cm^{3}}$.
Errors are $1\sigma$ unless stated otherwise. 
Labels correspond to the values represented in dark blue and with squared markers in figure \ref{fig:rhoDM-all-vertical}.
\label{Tab:rhoDM-values-rot-curve}}
\centering
\begin{tabular}{lcllcr}
\hline\\[-2ex]
\textbf{Label} & $\boldsymbol R$ \textbf{range} & \textbf{Tracer description} & $\boldsymbol \rhoDMlocal$ & $\boldsymbol R_\odot$ & \textbf{Ref.}\\
\hline\\[-2ex]
% Pato+15
Pato+15\footnote{The quoted $\rhoDMlocal$ corresponds to their preferred value. A wider $2\sigma$ range between $\sim \text{0.39--0.52}\,\mathrm{GeV/cm^3}$ can be read from their table II, depending on the configuration of the baryonic and dark matter profiles.}
& $\text{2.5--25}$
& \texttt{galkin} compilation
& $0.42\pm 0.03\,(2\sigma)$ 
& $8.0$
& \cite{Pato:2015dua} \\
\hline\\[-2ex]
% Benito+19
Benito+19
& $\text{2.5--22}$
& \texttt{galkin} compilation
& $\text{0.3--0.8}\, (2\sigma )$ 
& $\text{7.5--8.5}$
& \cite{Benito:2019ngh}
\\
\hline\\[-2ex]
% Benito+20
Benito+20
& $\text{2.5--22}$
& \texttt{galkin} compilation
& $\text{0.4--0.8}\, (2\sigma )$ 
& $8.178$
& \cite{Benito:2020lgu}
\\
\hline\\[-2ex]
% Karukes+19
Karukes+19\footnote{The quoted error for $\rhoDMlocal$ is only statistical. A systematic error of $\pm 0.01\,\mathrm{GeV/cm^3}$ should also be associated to this value, according to the authors.}
& $\text{2.5--100}$
& \texttt{galkin} compilation
& $0.43 \pm 0.02$ 
& $8.34$
& \cite{Karukes:2019jxv}
\\
& & PRCGs from \cite{Huang:1604.01216}\\
& & HKGs  from \cite{Huang:1604.01216}\\
\hline\\[-2ex]
% Huang+16
Huang+16
& $\text{4.6--100}$
& HI measurements
& $0.32 \pm 0.02$ 
& $8.34$
& \cite{Huang:1604.01216}
\\
& & $\sim 16\,000$ PRCGs \\
& & $\sim 5700$ HKGs \\
\hline\\[-2ex]
% Lin+19
Lin+19
& $\text{4.6--100}$
& same data as \cite{Huang:1604.01216}
& $0.51 \pm 0.09$ 
& $8.0$
& \cite{Lin:2019yux}
\\
\hline\\[-2ex]
% deSalas+19
deSalas+19\footnote{The quoted $\rhoDMlocal$ correspond to the result of the analyses fitting two different baryonic models: B1 and B2 (see section \ref{sec:rhoDM:rot-curve}).}
& $\text{5--25}$
& $v_{\rm c}(R)$ data from \cite{Eilers:1810.09466}
& $0.30\pm 0.03$ (B1) 
& $8.122\pm 0.031$
& \cite{deSalas:2019pee}
\\
& & & $0.38\pm 0.04$ (B2)
\\
\hline\\[-2ex]
% Sofue_20
Sofue\_20\footnote{The quoted $\rhoDMlocal$ value corresponds to the best-fitting NFW dark matter halo. A wider range of $\rhoDMlocal = \text{0.28--0.42}\,\mathrm{GeV/cm^3}$ is covered by the results of the analyses of other dark matter profiles.}
& $\text{1--100}$
& own compilation
& $0.36 \pm 0.02$ 
& $8.0$
& \cite{Sofue:2020rnl}
\\
\hline
\hline
\end{tabular}
\vspace{-2ex}                    % SHIFT FOOTNOTE UP
\end{minipage}
}
\end{figure}
%%%%%%%%%%%%%%%%%%%%%%%%%%%%%%%%%%%%%%%%%%%%%%%%%%%%%%%%
%%%
%%%           TABLE with ROT. CURVE (end)
%%%
%%%%%%%%%%%%%%%%%%%%%%%%%%%%%%%%%%%%%%%%%%%%%%%%%%%%%%%%

Inferences of $\rhoDMlocal$ from the Milky Way's circular velocity curve belong to the category of global estimates (see section~\ref{sec:spatial-coverage}). 
Other global studies are included in section \ref{sec:rhoDM:DF-global}. However, given the long list of papers that fitted the circular velocity curve to obtain $\rhoDMlocal$, we have separated the studies of the present section from section \ref{sec:rhoDM:DF-global}. Notice that in section \ref{sec:rhoDM:DF-global:RC} we include other studies that also used circular velocity data.
These works do not fit entirely into the category of a circular velocity curve method, because they also considered additional Galactic observations, which is why they have their own section.

For the studies presented here, there are systematic issues associated with the Galactocentric radial range of the data. As pointed out by \cite{Chemin:1504.01507}, some of the methods that measure circular velocities do not perform well for distances closer than $R \sim 4.5\,\mathrm{kpc}$ from the Galactic centre, partly because the gravitational potential this close to the centre strongly deviates from axisymmetry.
Additionally, one needs to be aware that the observed velocity is typically the azimuthal velocity of the tracer, not its circular velocity, and one needs to transform from one to the other.

Another important issue concerns the combination of data sets from different observations, which might have assumed different solar parameters, such as the specific value of $R_\odot$ or the Sun's peculiar velocity. These differences can add extra complications to the analyses as they should be considered and corrected for.
Moreover, different tracers might not have been equally affected by perturbations along the formation history of the Galaxy, and hence it is not guaranteed that their individual analyses result in the exact same gravitational potential under the assumptions of the Ideal Galaxy. Therefore, it could be problematic to combine several separate tracer populations when estimating $\rhoDMlocal$.
The discrepancies between the results of different tracer populations can inform us about the uncertainties of the assumed Galactic model, specifically in relation to the past history and current state of the Milky Way. 

The recent $\rhoDMlocal$ estimates that we are going to discuss in this section are summarised in table~\ref{Tab:rhoDM-values-rot-curve}.
Although most of the $\rhoDMlocal$ values shown in the table agree well with each other, some small tensions remain, which are mostly caused by different approaches when treating the baryonic content of the Milky Way.

The first four references followed a similar strategy, fitting to a circular velocity curve obtained from a compilation of observations. That choice was proven to be very useful when individual observations were still not very precise and the systematic uncertainties of fitting data from different tracers could be neglected. However, with the arrival of precise observations, such as those provided in ref.~\cite{Eilers:1810.09466}, this approach could lead to large systematic uncertainties.
References \cite{Huang:1604.01216} and \cite{Lin:2019yux} used the same data, but with different models for the baryonic components, which explains the difference in their quoted $\rhoDMlocal$ values. Reference \cite{deSalas:2019pee} fitted the data presented in ref.~\cite{Eilers:1810.09466} to several Galactic mass models, changing both the dark matter and baryonic distributions and varying the relevant parameters in each model. Finally, ref.~\cite{Sofue:2020rnl} presents a review of Galactic circular velocity observations, including a fit to a global circular velocity compilation within a wide but 
compromising range of $R=\text{1--100}\,\mathrm{kpc}$.

Starting with the works of refs.~\cite{Pato:2015dua,Benito:2019ngh,Karukes:2019jxv,Benito:2020lgu}, they share most of the observations included in their analyses.
In particular, both ref.~\cite{Pato:2015dua} and refs.~\cite{Benito:2019ngh,Benito:2020lgu} include the kinematics of 2174 gas, 506 star and 100 maser observations, that some of the authors of those papers collected into a single compilation named \texttt{galkin} \cite{Pato:2017yai}. This set of data covers a radial distance of $2.5 \,\mathrm{kpc} \lesssim R \lesssim 22 \,\mathrm{kpc}$.
An extended compilation was used in ref.~\cite{Karukes:2019jxv}, including data from ref.~\cite{Huang:1604.01216} to reach greater distances from the Galactic centre.

Reference \cite{Pato:2015dua} fitted both a generalised Navarro-Frenk-White (gNFW) and an Einasto spherical dark matter profile, together with a total of 70 different baryonic configurations that resulted from combining 7 different bulge shapes, 6 different stellar disc models and 2 gas disc profiles. 
Their preferred value, $\rhoDMlocal = (0.42 \pm 0.03)\,\mathrm{GeV/cm^3}$ ($2\sigma$ uncertainties), was obtained from a specific baryonic configuration, in an analysis that fixed $R_\odot = 8.0\,\mathrm{kpc}$ and $v_{\rm c} (R_\odot ) = 230\,\mathrm{km/s}$.
However, a broader $2\sigma$ range of $\rhoDMlocal = \text{0.39--0.52} \,\mathrm{GeV/cm^3}$ could be considered when varying among their baryonic configurations, as extracted from their table~II.

Reference~\cite{Benito:2019ngh} revisited the analysis of ref.~\cite{Pato:2015dua} using the same observations but a different methodology. In particular, contrary to the unbinned method applied in ref.~\cite{Pato:2015dua}, the authors of ref.~\cite{Benito:2019ngh} divided the observed $v_{\rm c}$ into 25 bins in $R$. They also fixed the peculiar azimuthal velocity of the Sun to $v_{\odot,\phi} = 12.25\,\mathrm{km/s}$ instead of fixing $v_{\mathrm{c}} (R_\odot)$. However, they allowed $R_\odot$ to change between $7.5\,\mathrm{kpc}$ and $8.5\,\mathrm{kpc}$. This wide range of $R_\odot$ values has a large impact in broadening the uncertainties of $\rhoDMlocal$. As a result of their analysis, they obtained a $2\sigma$ range of $\rhoDMlocal = \text{0.3--0.8} \,\mathrm{GeV/cm^3}$.
This paper has been recently updated in ref.~\cite{Benito:2020lgu}, where $R_\odot = 8.178\,\mathrm{kpc}$ was now fixed, but the value of $v_{\odot,\phi}$ was allowed to change instead.
This decision plays a similar role to varying $R_\odot$, explaining their still wide range of $\rhoDMlocal = \text{0.4--0.8} \,\mathrm{GeV/cm^3}$ at $2\sigma$.
The value of $v_{\odot,\phi}$ is adapted from a range in $v_{\mathrm{c}} (R_\odot) = \text{218--240}\,\mathrm{km/s}$ taken from ref.~\cite{Eilers:1810.09466}.
The circular velocity data, however, comprises the same \texttt{galkin} compilation as in the previous paper, without the addition of all the precise data---covering $R = \text{5--25}\,\mathrm{kpc}$---available in ref.~\cite{Eilers:1810.09466}.

Reference \cite{Karukes:2019jxv} also revisited the analysis of ref.~\cite{Pato:2015dua}. 
They fixed $R_\odot = 8.34\,\mathrm{kpc}$ and $v_{\rm c}(R_\odot) = 239.89\,\mathrm{km/s}$, but their radial coverage was increased up to $R \sim 100\,\mathrm{kpc}$ by adding data from ref.~\cite{Huang:1604.01216}. 
They pointed out that some of the 25 data samples from the \texttt{galkin} compilation are not compatible with each other. 
For this reason, the authors only included 12 data samples in their analysis, which were chosen based on a Bayesian evidence comparison with the rest of the \texttt{galkin} data set, starting with the measurements from ref.~\cite{Malhotra:1994qj}.
They quoted a value of $\rhoDMlocal = [0.43 \pm 0.02\,(\mathrm{stat.})\pm 0.01\,(\mathrm{syst.}) ] \,\mathrm{GeV/cm^3}$ as a result of their analysis, which they found to be robust under changes in the baryonic distributions.

The analysis of ref.~\cite{Huang:1604.01216} studied the circular velocity curve for a wide radial range of $4.6 \,\mathrm{kpc} < R \lesssim 100\,\mathrm{kpc}$. The authors derived the circular velocities at $R \gtrsim R_\odot$ from $\sim 5700$ halo K giant (HKG) stars and $\sim 16\,000$ primary red clump giant (PRCG) stars, while for the inner region $R \lesssim R_\odot$ they used the HI measurements of ref.~\cite{Fich:1989}.
In order to infer $\rhoDMlocal$, they fitted the circular velocity data to a model consisting of three exponential discs, from which they only varied the scale length of the thin and thick stellar discs; an exponential bulge with fixed shape; an NFW dark matter halo, and the presence of two hypothetical caustic dark rings. 
Out of all the references included in this report, only ref.~\cite{Huang:1604.01216} considered such caustic rings.
Although they provide a nice fit to some features in their circular velocity curve at $R \sim 10\,\mathrm{kpc}$ and $R \sim 20\,\mathrm{kpc}$, these features are hardly visible in the curves derived from Gaia DR2 observations \cite{Eilers:1810.09466,Mroz:1810.02131}. 
As a result of their work, the authors of ref.~\cite{Huang:1604.01216} obtained a value of $\rhoDMlocal = (0.32\pm 0.02)\,\mathrm{GeV/cm^3}$.

The authors of ref.~\cite{Lin:2019yux} analysed the same data as ref.~\cite{Huang:1604.01216}, but they followed a different strategy. Inspired by the work of ref.~\cite{Pato:2015dua}, they fitted the data to a wide collection of Galactic mass models, which were built from 56 fixed baryonic configurations and four different 2-parameter dark matter profiles. 
They found that, for most of their baryonic configurations, a dark matter NFW profile provides a better fit to the data. For their best-fitting configuration---based on a $\chi^2$ comparison---they quoted an estimated $\rhoDMlocal = (0.51\pm 0.09)\, \mathrm{GeV/cm^3}$. 
However, we should keep in mind that the quoted uncertainty comes from a fit with a fixed configuration of baryons, even if its $\chi^2$ is smaller than for the other configurations. Including the uncertainty due to the modelling of the baryonic components should bring the result from refs.~\cite{Huang:1604.01216,Lin:2019yux} in closer agreement.

The authors of ref.~\cite{deSalas:2019pee} estimated the value of $\rhoDMlocal$ from the accurate $v_{\rm c}$ data presented in ref.~\cite{Eilers:1810.09466}, which are based on a study of spectrophotometric parallaxes for red giant stars, obtained from a cross match of Gaia DR2, 2MASS, WISE and APOGEE observations \cite{Hogg:1810.09468}. 
The data was derived assuming a solar azimuthal velocity of $245.8\,\mathrm{km/s}$ in the Galactocentric frame, and covers a range of $R = \text{5--25}\,\mathrm{kpc}$. 
Reference \cite{deSalas:2019pee} analysed this data for different Galactic mass models, where the parameters of both the baryonic and dark matter profiles were allowed to change. The resulting $\rhoDMlocal$ was found to depend more strongly on the assumed shape for the baryonic components than on the specific shape of the dark matter halo, even if the parameters of both components were allowed to vary. Despite the unprecedented precision of the circular velocity data from ref.~\cite{Eilers:1810.09466}, the different analysed configurations in ref.~\cite{deSalas:2019pee} covered a $1\sigma$ range of $\rhoDMlocal = \text{0.27--0.42}\,\mathrm{GeV/cm^3}$, with the leading uncertainties coming from the lack of knowledge on the distribution of baryons.
In particular, they fitted three different shapes for the dark matter halo together with two baryonic models based on observations: one model (tagged B1) consisted of two Miyamoto-Nagai discs and a Plummer bulge (following the shapes chosen in ref.~\cite{Pouliasis:1611.07979}), and the other model (tagged B2) was composed of several double-exponential discs and a Hernquist bulge (following the model of ref.~\cite{Misiriotis:2006qq}). The independent analysis of the two baryonic models gave $\rhoDMlocal \simeq (0.30 \pm 0.03) \,\mathrm{GeV/cm^3}$ for the model B1, and $\rhoDMlocal \simeq (0.38 \pm 0.04) \,\mathrm{GeV/cm^3}$ for the model B2, with very little dependence on the chosen shape for the dark matter halo.

The last work presented in table~\ref{Tab:rhoDM-values-rot-curve}, ref.~\cite{Sofue:2020rnl}, corresponds to a review of Milky Way rotation curve estimates, from which a combined rotation curve was built up to $R \sim 100\,\mathrm{kpc}$.
This reference also includes the results from a fit to the Galactic circular velocity between $R = \text{1--100}\,\mathrm{kpc}$, assuming a de Vaucouleurs bulge and a razor-thin disc as representative of the Galactic baryonic components. 
The best-fitting halo of this study---an NFW---resulted in $\rhoDMlocal = (0.36 \pm 0.02)\,\mathrm{GeV/cm^3}$, although a wider range of $\rhoDMlocal = \text{0.28--0.42} \,\mathrm{GeV/cm^3}$ is covered by the results of the analyses when fitting the data to several cored dark matter profiles, as shown in their table~3.
However, the fit to the data was found to be much worse for the cored profiles than for the NFW case.
One thing to note is that the study of ref.~\cite{Sofue:2020rnl} assumed $R_\odot = 8.0\,\mathrm{kpc}$ and $v_{\rm c}(R_\odot) = 238\,\mathrm{km/s}$ to be fixed. As the authors mention, varying these quantities over $|\Delta R|\sim 0.1 \,\mathrm{kpc}$ and $|\Delta v_{\rm c}(R_\odot) | \sim 10\,\mathrm{km/s}$ could induce a systematic change in the quoted value of $\rhoDMlocal$ up to $\sim 0.1 \,\mathrm{GeV/cm^3}$, in line with the results presented by the other studies included in table~\ref{Tab:rhoDM-values-rot-curve}.

%%%%%%%%%%%%%%%%%%%%%%%%%%%%%%%%%%%%%%%%%%%%%%%%%%%%%%%%%%%%%%%%%%%%%%
\subsection{From global mass modelling of the Milky Way}\label{sec:rhoDM:DF-global}

The works that we discuss in this section have applied a wide variety of methods. However, all of them share the pursuit of fitting a global mass model of the Milky Way, and most of them consider different types of observations. 
One advantage of considering different types of observations is that it is easier to fit a global mass model for the Galaxy, since each observation can be sensitive to a different parameter of the model. However, one must also pay special attention to the compatibility of the observations, which might induce additional systematic uncertainties if they are not carefully included in the fit.

The studies included in the present section, summarised in table~\ref{Tab:rhoDM-values-global}, are divided into four categories, which are ordered according to the volume covered by their main observations. The first category includes works that were based on a distribution function fitting of local disc stars (section \ref{sec:rhoDM:DF-global:DF}); the second category presents a Jeans anisotropic modelling of close giant disc stars (section \ref{sec:rhoDM:JAM-disc}); the third category includes studies whose main tracers constrained circular velocities (section \ref{sec:rhoDM:DF-global:RC}), and the fourth category presents studies that analysed halo stars (section \ref{sec:rhoDM:DF-global:halo}).
All studies included in table~\ref{Tab:rhoDM-values-global} can be understood as global studies (see the discussion in section~\ref{sec:spatial-coverage}), but the analysis method of each category makes them very different from each other. In the next subsections we are going to describe the main aspects of each of these works.

%%%%%%%%%%%%%%%%%%%%%%%%%%%%%%%%%%%%%%%%%%%%%%%%%%%%%%%%
%%%
%%%           TABLE with GLOBAL (beginning)
%%%
%%%%%%%%%%%%%%%%%%%%%%%%%%%%%%%%%%%%%%%%%%%%%%%%%%%%%%%%
\begin{figure}
\noindent{\footnotesize
\begin{minipage}{\textwidth}   % GUESS (OR CALCULATE) MINIPAGE WIDTH
\renewcommand\footnoterule{}     % ELIMINATE LITTLE LINE SEPARATER
\captionof{table}{Recent $\rhoDMlocal$ estimates from different studies based on fitting a global Galactic mass model (discussed in section~\ref{sec:rhoDM:DF-global}).
Distances are expressed in units of $\mathrm{kpc}$ and $\rhoDMlocal$ in units of $\mathrm{GeV/cm^{3}}$.
Primed quantities correspond to a reference system centred at the location of the Sun.
Errors are $1\sigma$ unless stated otherwise.
Labels correspond to the values presented in figure \ref{fig:rhoDM-all-vertical}.
\label{Tab:rhoDM-values-global}}
\centering
\begin{tabular}{lcllcr}
\hline\\[-2ex]
\multicolumn{4}{l}{\qquad Distribution function fitting analyses of disc stars}\\
\hline\\[-2ex]
\textbf{Label} & \textbf{Main cuts} & \textbf{Observational inputs} & $\boldsymbol\rhoDMlocal$ & $\boldsymbol R_\odot$ & \textbf{Ref.}\\
\hline\\[-2ex]
% Bienayme+14
Bienaym\'e+14
& $\Delta R' < 0.5$
& 4600 red clump stars
& $0.542\pm 0.042$ 
& $8.5$
& \cite{Bienayme:2014kva}
\\
& $0.2< -z < 2$ & ($\text{RAVE}\times \text{2MASS} \times \text{UCAC}$) \\
\hline\\[-2ex]
% Piffl+14/15
Piffl+14/15\footnote{The quoted value corresponds to the range of minor-major axis ratios $q=\text{0.75--0.95}$ found in \cite{Piffl:2015xua} applied to eq.~(22) in \cite{Piffl:2014mfa}.}\footref{footnote:DF-fitting}
& $|R-R_\odot| < 1$
& \textbullet\ Giant star kinematics (RAVE)
& $\text{0.50--0.62}$ 
& $8.3$
& \cite{Piffl:2014mfa,Piffl:2015xua}
\\
\cline{1-1}\cline{4-6}
& $|z|< 1.5$ 
& \phantom{\textbullet} ($|R-R_\odot| < 1$ \& $|z|< 1.5$)
\\
% Binney+15
Binney+15\footnote{The quoted value corresponds to their best-fit model.}\footref{footnote:DF-fitting}
&
& \textbullet\ Vertical dens. distr. (SDSS)
& $0.50$ 
& $8.3$
& \cite{Binney:2015gaa}
\\
\cline{1-1}\cline{4-6}
& & \textbullet\ Gas terminal velocities
\\
Cole+17\footnote{\label{footnote:DF-fitting}References \cite{Piffl:2014mfa,Piffl:2015xua,Binney:2015gaa,Cole:2016gzv} carried out very similar analyses that improved upon the previous one. The listed spatial cuts apply to the tracer giant stars from RAVE.}
& 
& \textbullet\ Maser kinematics
& $\text{0.456--0.527}$ 
& $8.3$
& \cite{Cole:2016gzv}
\\
\cline{1-1}\cline{4-6}
& 
& \textbullet\ Proper motion of SgrA*
\\
\hline\\[-2ex]
\multicolumn{4}{l}{\qquad Jeans anisotropic modelling of disc stars }\\
\hline\\[-2ex]
\textbf{Label} & \textbf{Main cuts} & \textbf{Tracer description} & $\boldsymbol\rhoDMlocal$ & $\boldsymbol R_\odot$ & \textbf{Ref.}\\
\hline\\[-2ex]
% Nitschai+20
Nitschai+20
& $3.6 < R < 12$
& $\sim 2$ million giant stars
& $0.437 \pm 0.076$
& $8.2$
& \cite{Nitschai:1909.05269}
\\
& $|z| < 2.5$
& (Gaia DR2)
\\
\hline\\[-2ex]
\multicolumn{4}{l}{\qquad Galactic mass models driven by circular velocity data}\\
\hline\\[-2ex]
\textbf{Label} & \textbf{Main cuts} & \textbf{Observational inputs} & $\boldsymbol\rhoDMlocal$ & $\boldsymbol R_\odot$ & \textbf{Ref.}\\
\hline\\[-2ex]
% McMillan_17
McMillan\_17
& $ 4 \lesssim R \lesssim 20$
& \textbullet\ Maser kinematics
& $0.40 \pm 0.04$ 
& $8.20\pm 0.09$
& \cite{McMillan:1608.00971}
\\
& & \textbullet\ Gas terminal velocities \\
& & \textbullet\ Proper motion of SgrA* \\
& & \textbullet\ Vertical force at $1.1\,\mathrm{kpc}$ \\
& & \textbullet\ Total mass at $50\,\mathrm{kpc}$ \\
\hline\\[-2ex]
% Cautun+20
Cautun+20
& $5 < R < 25$
& \textbullet\ $v_{\rm c}(R)$ curve from \cite{Eilers:1810.09466}
& $0.33 \pm 0.02$ 
& $8.12 \pm 0.03$
& \cite{Cautun:2019eaf}
\\
& 
& \phantom{\textbullet\ }($\text{Gaia DR2}\times \text{APOGEE}$
\\
&
& \phantom{\textbullet\ }$\times \text{WISE}\times \text{2MASS}$)
\\
&
& \textbullet\ Total Galactic mass
\\
& & \textbullet\ Vertical force at $1.1\,\mathrm{kpc}$ \\
\hline\\[-2ex]
\multicolumn{4}{l}{\qquad Galactic mass models from halo stars}\\
\hline\\[-2ex]
\textbf{Label} & \textbf{Main cuts} & \textbf{Observational inputs} & $\boldsymbol\rhoDMlocal$ & $\boldsymbol R_\odot$ & \textbf{Ref.}\\
\hline\\[-2ex]
% Wegg+19
Wegg+19\footnote{The quoted $\rhoDMlocal$ corresponds to their fidutial (conservative) fit to an ellipsoidal version of an Einasto dark matter halo, for which a minor-major axis ratio of $q = 1.00\pm 0.09$ was found.}
& $1.5 < r < 20$
& $15\,651$ RR Lyrae halo stars
& $0.35 \pm 0.08$ 
& $8.2$
& \cite{Wegg_2019}
\\
& & ($\text{PanSTARRS1} \times \text{Gaia DR2}$)
\\
\hline\\[-2ex]
% Hattori+20
Hattori+20
& $r' < 20$
& \textbullet\ $16\,197$ RR Lyrae halo stars
& $0.342 \pm 0.007$
& $8.178$
& \cite{Hattori:2012.03908}
\\
& & \phantom{\textbullet\ }(Gaia DR2)
\\
& & \textbullet\ $v_{\rm c}(R)$ curve from \cite{Eilers:1810.09466}
\\
& & \textbullet\ Vertical force at $1.1\,\mathrm{kpc}$
\\
\hline
\hline
\end{tabular}
\vspace{-2ex}                    % SHIFT FOOTNOTE UP
\end{minipage}
}
\end{figure}
%%%%%%%%%%%%%%%%%%%%%%%%%%%%%%%%%%%%%%%%%%%%%%%%%%%%%%%%
%%%
%%%           TABLE with GLOBAL (end)
%%%
%%%%%%%%%%%%%%%%%%%%%%%%%%%%%%%%%%%%%%%%%%%%%%%%%%%%%%%%

\subsubsection{Global mass models from a distribution function fitting of disc stars}\label{sec:rhoDM:DF-global:DF}

The works shown in the first part of table~\ref{Tab:rhoDM-values-global} analysed disc stars under a distribution function fitting approach (see section \ref{sec:method:DF-modelling}), with the goal of providing global Galactic mass models compatible with the observations. Given the numerical complexity of this method, not all references in the table included uncertainties for the quoted $\rhoDMlocal$.
The obtained values agree well with most of the local analyses referenced in the first half of table~\ref{Tab:rhoDM-values-local}. Since the analyses of the first half of both tables \ref{Tab:rhoDM-values-local} and \ref{Tab:rhoDM-values-global} considered local stars as their main tracers, the agreement in their results indicates that the obtained $\rhoDMlocal$ does not depend too strongly on the specific method when similar data is analysed, at least when similar assumptions are also made regarding the current status of the Galaxy.

Let us now describe each study individually.

Starting with ref.~\cite{Bienayme:2014kva}, the authors studied a population of about 4600 red clump stars, whose properties were obtained from a cross match of observations from RAVE, 2MASS and UCAC3, contained in a cylinder centred at the Sun's position and defined by $\Delta R < 0.5\,\mathrm{kpc}$ and $z = \text{0.2--2}\,\mathrm{kpc}$ towards the Galactic south pole. The stars are divided into three metallicity subsamples. This data set is complemented with $\sim 300$ red clump stars from Hipparcos and Elodie at $z = \text{0--0.8}\,\mathrm{kpc}$ to the Galactic north. 
The estimated $\rhoDMlocal$ is derived from a fit of a simple two-component Galactic model to the observed vertical force $K_z(z)$.
Assuming that all relevant baryonic mass is enclosed in a fixed double-exponential disc, ref.~\cite{Bienayme:2014kva} fitted a spherical dark matter halo, imposing a flat rotation curve close to $R_\odot$, which they fix to a value of $8.5 \,\mathrm{kpc}$. Although they mention that the resulting $\rhoDMlocal$ is not too sensitive to a change in the parameters of their Galactic disc, the value they obtained for $v_{\rm c}(R_\odot) = 267\, \mathrm{km/s}$ is too large when compared to observations. They suggested several possibilities to reduce this tension, among them the consideration of a cored or flattened shape for the dark matter halo.

The works of \cite{Piffl:2014mfa,Piffl:2015xua,Binney:2015gaa,Cole:2016gzv} carried out very similar analyses with successive improvements.
The distribution function fitted in these studies was expressed in terms of action-angle variables, and applied to $\sim 180\,000$ RAVE giant stars within the Galactocentric cylindrical shell defined by $|R - R_\odot | < 1\,\mathrm{kpc}$ and $|z| < 1.5 \,\mathrm{kpc}$.
Apart from their main tracer stars, these studies included other information to constrain their Galactic models, such as the vertical distribution of local stars within $|z| \sim 3 \,\mathrm{kpc}$ \cite{Juric:2005zr}, gas terminal velocities or maser observations.
Two key aspects on the dark matter halo distribution function developed in this sequence of papers are: a) the addition in ref.~\cite{Piffl:2015xua} of adiabatic contraction when baryons are present, and b) the inclusion of the possibility of a core in the inner part of the halo \cite{Cole:2016gzv}.
The latter made the fitted models from the previous ref.~\cite{Binney:2015gaa} compatible with microlensing data. 

\subsubsection{Jeans anisotropic modelling of disc stars}\label{sec:rhoDM:JAM-disc}

Another global study that analysed more or less local observations is the work presented in ref.~\cite{Nitschai:1909.05269}. The baryonic model of this work, which is based on the model from ref.~\cite{McMillan:1608.00971}, is almost entirely fixed. 
Therefore, the analysis focused on inferring the properties of a global Galactic dark matter halo. For this purpose, the authors studied the 6-dimensional phase space information of almost 2 million giant stars from Gaia DR2. The data was selected from a volume defined by the following Galactocentric cylindrical cuts: $3.6\,\mathrm{kpc} \lesssim R \lesssim 12\,\mathrm{kpc}$; $|z| \lesssim 2.5\,\mathrm{kpc}$; and $|\phi| \lesssim 15^\circ$; with additional selection cuts in absolute magnitude and intrinsic colour. 
The analysis applied the Jeans anisotropic method from \cite{Cappellari:1907.09894}, assuming axisymmetry.
The method consists in fitting the azimuthal velocity $v_\phi$ and the three velocity dispersions from the observations, which are binned in $(R,z)$, to a spherically aligned velocity ellipsoid.
The result of the analysis gave $\rhoDMlocal = (0.437 \pm 0.076)\,\mathrm{GeV/cm^3}$.

\subsubsection{Global mass models driven by circular velocity data}\label{sec:rhoDM:DF-global:RC}

The studies shown in the third part of table \ref{Tab:rhoDM-values-global} benefit considerably from circular velocity data. However, unlike the works presented in section~\ref{sec:rhoDM:rot-curve}, they include other Galactic observations in a more varied and complete way.
Their results for $\rhoDMlocal$ are in good agreement with those quoted in table \ref{Tab:rhoDM-values-rot-curve}, which generally show a smaller central value than the local studies of table~\ref{Tab:rhoDM-values-local}. 

The first work included in the third category of table \ref{Tab:rhoDM-values-global} is that of ref.~\cite{McMillan:1608.00971}. The authors of this analysis considered different observations, such as circular velocities inferred from masers and terminal velocities; the vertical force at $z = 1.1\,\mathrm{kpc}$ from ref.~\cite{Kuijken:1991}, or a constraint to the total mass of the Galaxy within $50 \,\mathrm{kpc}$ from its centre.
A spherical gNFW dark matter profile is assumed, together with axisymmetric shapes for the baryonic components---bulge, stellar and gaseous discs---whose parameters are fitted using well motivated prior distributions.
The inferred Galactic model gives that $\rhoDMlocal = (0.40\pm 0.04)\, \mathrm{GeV/cm^3}$.

The second work of this category corresponds to ref.~\cite{Cautun:2019eaf}.
The most constraining observation used in this analysis is the circular velocity curve measured in ref.~\cite{Eilers:1810.09466}, to which they added other constraints such as the total Galactic mass estimated in ref.~\cite{Callingham:1808.10456}, and the vertical force at $z = 1.1\,\mathrm{kpc}$ from ref.~\cite{Kuijken:1991}.
The main difference in their study with respect to ref.~\cite{McMillan:1608.00971} and other studies based on the $v_{\rm c}(R)$ data from ref.~\cite{Eilers:1810.09466} (e.g., \cite{deSalas:2019pee}) is that they focused on the contraction of the dark matter halo due to the condensation of baryons, advancing in a similar direction as \cite{Binney:2015gaa,Cole:2016gzv}, but with a completely different methodology. Comparing between several Milky Way-like simulated galaxies and their dark matter only counterparts, the authors of ref.~\cite{Cautun:2019eaf} found a relation to account for the dark matter halo contraction, which they applied as part of their Galactic model---whose baryonic components are an axisymmetric bulge, two double exponential stellar discs, two gas discs and a halo accounting for the circumgalactic medium. As a result of their analysis, they found $\rhoDMlocal = (0.33 \pm 0.02)\,\mathrm{GeV/cm^3}$.

\subsubsection{Global mass models from halo stars} \label{sec:rhoDM:DF-global:halo}

Most of the studies listed in table \ref{Tab:rhoDM-values-global} used observations that either belong to the disc or are reasonably close to the Galactic plane.
However, properties of the Milky Way halo, including $\rhoDMlocal$, can also been inferred from the kinematics of halo objects. 
The result of these studies serve as a complementary way of inferring $\rhoDMlocal$, since they constitute truly global estimates.

We include two studies of RR Lyrae halo stars in the last part of table~\ref{Tab:rhoDM-values-global}.
Reference~\cite{Wegg_2019} studied more than $15\,000$ RR Lyrae halo stars, which are widely distributed over the Galaxy, from Pan-STARRS1 \cite{Sesar:1611.08596} cross matched with Gaia DR2. 
Following azimuthally average spherical Jeans equations, they fitted several shapes for the dark matter halo to the kinematics of their RR Lyrae sample. Their Galactic mass model was complemented with a baryonic potential inspired in the model used in ref.~\cite{Portail:1502.00633}. As a result, the authors of ref.~\cite{Wegg_2019} quoted a conservative value of $\rhoDMlocal = (0.35 \pm 0.08) \,\mathrm{GeV/cm^3}$ when they fitted an Einasto dark matter halo. The result of fitting other shapes gave a similar central value, but with about half the uncertainty of the result from the Einasto profile (see table 1 of ref.~\cite{Wegg_2019}). In addition, they allowed all dark matter halo shapes to have a free minor-major axis ratio $q$, keeping the assumption of axisymmetry. All their fits gave values of $q \approx 1$.

The second study shown in the last part of table \ref{Tab:rhoDM-values-global}, carried out in ref.~\cite{Hattori:2012.03908}, focused on a distribution function fitting of $\sim 16\,000$ RR Lyrae halo stars extracted from a larger sample from ref.~\cite{Iorio:1808.04370}. The Galactic mass model assumed in ref.~\cite{Hattori:2012.03908} is based on the model of ref.~\cite{McMillan:1608.00971}, with the extra consideration that the gNFW dark matter halo is allowed to be oblate. In addition to the main RR Lyrae sample, a constraint in both the circular velocity curve from \cite{Eilers:1810.09466} and the vertical force $K_{z}$ at $z = 1.1\,\mathrm{kpc}$ from \cite{Bovy:2013raa} are included in the analysis. As a result, the authors of this work found $\rhoDMlocal = (0.342 \pm 0.007)\,\mathrm{GeV/cm^3}$.

%%%%%%%%%%%%%%%%%%%%%%%%%%%%%%%%%%%%%%%%%%%%%%%%%%%%%%%%%%%%%%%%%%%%%%
\subsection{Summary of $\rhoDMlocal$ estimates}\label{sec:rhoDM:summary}

In the previous sections we have presented the most recent estimates of $\rhoDMlocal$, which have been obtained from a large variety of methods. 
All the results are depicted in figure~\ref{fig:rhoDM-all-vertical}, with the exception of those discussed in section \ref{sec:rhoDM:local-and-very-local:very-local}, which were excluded because of their large uncertainties. 
It is notable that, although deviations exist between estimates, all values shown in the figure are compatible with $\rhoDMlocal \gtrsim 0.3\,\mathrm{GeV/cm^3}$, and most of them with $\rhoDMlocal \lesssim 0.6\,\mathrm{GeV/cm^3}$.

Starting with the estimates that are local (section \ref{sec:rhoDM:local-and-very-local:local}), represented at the top of figure~\ref{fig:rhoDM-all-vertical} in brown colour and with triangular markers, the most recent analyses seem to favour values of $\rhoDMlocal \sim \text{0.4--0.6} \,\mathrm{GeV/cm^3}$, with the exceptions of \cite{Hagen:1802.09291} and the Galactic south analysis of ref.~\cite{Salomon:2020eer}. As discussed in some of these references, the wide range of values is possibly reflecting a perturbed local environment, especially towards the Galactic north \cite{Salomon:2020eer}. 

We do not include in figure~\ref{fig:rhoDM-all-vertical} the last two works of table \ref{Tab:rhoDM-values-local} (section~\ref{sec:rhoDM:local-and-very-local:very-local}), which have attempted to provide a $\rhoDMlocal$ estimate from very local observations---within $\sim 200\,\mathrm{pc}$ from the Sun. Although these estimates truly represent the value of $\rhoDMlocal$ at the Sun's position, the uncertainty of the results is too large to be informative. Large uncertainties are expected in this type of analyses given the dominant contribution from baryons to the vertical gravitational potential this close to the Sun's location. 
Therefore, these measurements are very sensitive to any systematic errors in the baryonic distribution, which are not unlikely. Furthermore, the results of ref.~\cite{Widmark:2020vqi} are indicative of time-varying dynamical effects, possibly in the form of a breathing mode, which would bias these very local estimates even further. Moreover, discrepancies are also apparent between the results of different stellar tracer populations, even within the same studies \cite{Schutz:2017tfp,Buch:2018qdr,Widmark:2018ylf}. We further comment on these issues in section \ref{sec:breaking-of-Ideal-Galaxy}.

\begin{figure}\label{fig:rhoDM-all-vertical}
\begin{center}
\includegraphics[width=0.5\textwidth]{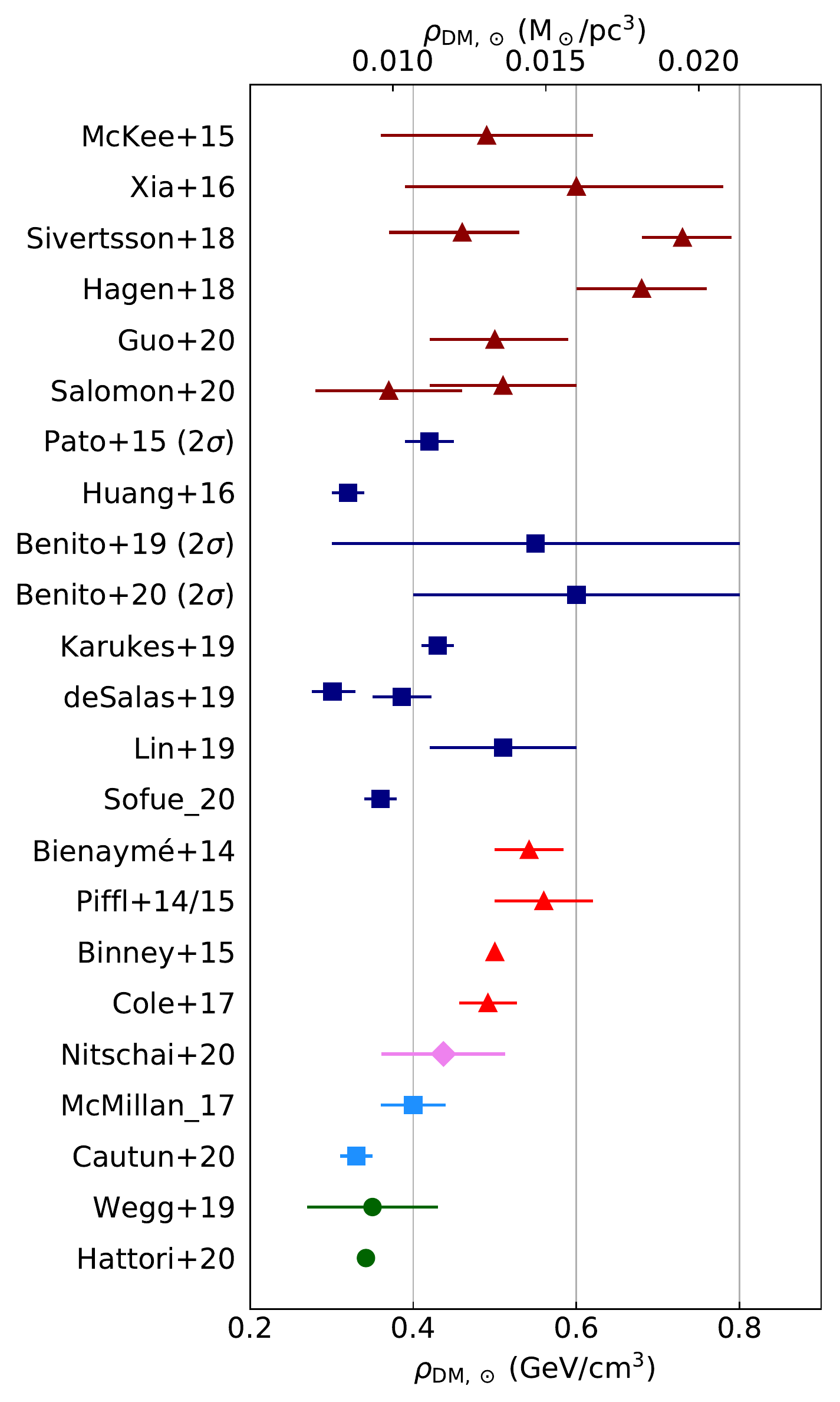}
\caption{Summary of recent $\rhoDMlocal$ estimates. The marker type indicates the main observation of the analyses: triangles for local observations, squares for circular velocities, a diamond for disc stars in an extended local region, and circles for halo stars. From top to bottom: the brown triangles correspond to the local studies presented in section \ref{sec:rhoDM:local-and-very-local:local}; the dark blue squares to the circular velocity analyses from section \ref{sec:rhoDM:rot-curve}; the red triangles to the Galactic mass models based on local observations, discussed in section \ref{sec:rhoDM:DF-global:DF}; the pink diamond to the Jeans anisotropic modelling of disc stars presented in section \ref{sec:rhoDM:JAM-disc}; the cyan squares to the circular-velocity-based Galactic mass models included in section \ref{sec:rhoDM:DF-global:RC}; and the green circles to the analyses of halo stars from section \ref{sec:rhoDM:DF-global:halo}. We do not include the very local analyses from section \ref{sec:rhoDM:local-and-very-local:very-local} because of their large error bars.}\label{fig:rhoDM-all-vertical}
\end{center}
\end{figure}

Global analyses based on the circular velocity curve of the Milky Way (section \ref{sec:rhoDM:rot-curve}) are included in dark blue at the middle of figure~\ref{fig:rhoDM-all-vertical}, marked with squared symbols. These studies tend to favour slightly smaller values of $\rhoDMlocal$ than local studies, with uncertainties that are generally smaller. Most of the estimates are well contained within a range of $\rhoDMlocal \sim \text{0.3--0.5}\,\mathrm{GeV/cm^3}$.
The assumed distribution of baryons in this type of studies plays as important a role as in local studies. Thanks to the observations of Gaia DR2, very precise estimates of $v_{\rm c}(R)$ within $R \sim \text{5--25}\,\mathrm{kpc}$ are currently available \cite{Eilers:1810.09466}. However, without a correspondingly precise knowledge of how baryons are distributed, it is not possible to disentangle the contribution to $v_{\rm c}(R)$ from baryons and dark matter. Therefore, the uncertainty of the resulting $\rhoDMlocal$ is dominated by the uncertainties in the baryonic distribution (e.g., \cite{deSalas:2019pee}). 

The results of recent global mass models (section~\ref{sec:rhoDM:DF-global}) are also included in figure~\ref{fig:rhoDM-all-vertical}.
Some studies focused on fitting the distribution function of disc stars---section~\ref{sec:rhoDM:DF-global:DF} and estimates in figure~\ref{fig:rhoDM-all-vertical} shown in red with triangular markers---complementing their analyses with other Galactic constraints. The results of these works prefer a value of $\rhoDMlocal \sim 0.5 \,\mathrm{GeV/cm^3}$, in better agreement with the outcome of the local analyses of section \ref{sec:rhoDM:local-and-very-local:local} than with the result of the circular velocity analyses of section \ref{sec:rhoDM:rot-curve}. In fact, the preferred $\rhoDMlocal$ value from the mass model analyses based on a distribution function fitting of disc stars agree very well to a similar analysis carried out in \cite{Salomon:2020eer} as a complementary check. However, as stated in that reference, this estimate could be biased because of a perturbed environment, especially towards the Galactic north. Indeed, when it is only stars from the Galactic south that are analysed in ref.~\cite{Salomon:2020eer}---for their main Jeans' analysis---a smaller value of $\rhoDMlocal = (0.37 \pm 0.09)\,\mathrm{GeV/cm^3}$ is found.

The result of ref.~\cite{Nitschai:1909.05269} is also included in figure \ref{fig:rhoDM-all-vertical}, with a diamond symbol and in pink colour. 
Their $\rhoDMlocal$ is compatible with both local and global estimates, which could be connected with the fact that their observations cover a volume that is in between the typical radial extension of local and global studies.

The two estimates towards the bottom of figure \ref{fig:rhoDM-all-vertical} that are shown in cyan with squared markers correspond to global mass models where the results are driven by the Milky Way's circular velocity curve (section~\ref{sec:rhoDM:DF-global:RC}). The inferred $\rhoDMlocal$ from both references are in agreement with the range of $\rhoDMlocal \sim \text{0.3--0.5}\,\mathrm{GeV/cm^3}$ compatible with most of the analyses from section \ref{sec:rhoDM:rot-curve}. In particular, ref.~\cite{Cautun:2019eaf} analysed the Gaia-DR2-inferred data from \cite{Eilers:1810.09466}---complemented with other Galactic constraints---finding a value compatible with the analyses of ref.~\cite{deSalas:2019pee}, which analysed the same $v_{\rm c}(R)$ data.

Finally, two analyses of halo stars (section~\ref{sec:rhoDM:DF-global:halo}) are also included at the bottom of figure~\ref{fig:rhoDM-all-vertical}, shown in green with circular markers.
The estimated $\rhoDMlocal$ of these two studies are likewise compatible with the global estimates based on the circular velocity curve. In addition, since these two references analysed halo stars, their results are highly complementary to the other studies, in particular to those based on local observations.

From the results shown in figure \ref{fig:rhoDM-all-vertical}, it seems clear that, in order to reduce their differences and aim at a robust estimate of $\rhoDMlocal$, it is very important to improve our Galactic models. In that sense, the natural next step should be to diagnose and reduce the systematic uncertainties, with the goal of achieving both a good precision and accuracy in the inferred $\rhoDMlocal$.

%%%%%%%%%%%%%%%%%%%%%%%%%%%%%%%%%%%%%%%%%%%%%%%%%%%%%%%%%%%%%%%%%%%%%%
%%%%%%%%%%%%%%%%%%%%%%%%%%%%%%%%%%%%%%%%%%%%%%%%%%%%%%%%%%%%%%%%%%%%%%
\section{The breaking of the Ideal Galaxy}\label{sec:breaking-of-Ideal-Galaxy}

There are several sources of uncertainties when estimating $\rhoDMlocal$.
Like for any other data-based study, observational uncertainties are unavoidable, but the precision of recent catalogues can lead to very precise $\rhoDMlocal$ estimates when rather strong assumptions are made regarding the current Galactic state and shape.
Therefore, the challenge when estimating $\rhoDMlocal$ is less a matter of precision and more about accuracy.
In that sense, it has become increasingly important to carefully treat the uncertainties associated with the Galactic modelling itself.
A too simplistic study can lead to very precise but inaccurate results, biasing the conclusions.
Fortunately, current astrophysical observations have reached sufficient precision for us to advance in this direction, and several studies have already initiated the journey.

In this section, we first discuss the origin of the differences between $\rhoDMlocal$ estimates (section~\ref{sec:diff-between-estimates}). Next, we comment on the importance of a steady-state and symmetric assumptions (sections \ref{sec:disequilibria} and \ref{sec:importance-symmetry}), followed by a discussion on the baryonic models (section \ref{sec:importance-baryons}). 
We finish the section by covering the hypothetical effect on $\rhoDMlocal$ from unconventional dark matter distributions (section \ref{sec:new-physics}).

%%%%%%%%%%%%%%%%%%%%%%%%%%%%%%%%%%%%%%%%%%%%%%%%%%%%%%%%%%%%%%%%%%%%%%
\subsection{Differences between recent estimates of \rhoDMlocal}\label{sec:diff-between-estimates}

The differences between categories of recent estimates have already been partially covered in section~\ref{sec:rhoDM:summary}. Here we are going to delve into the origin of these differences. 

At a first glance, figure~\ref{fig:rhoDM-all-vertical} shows that the estimates based on global observations---those with squared or circular markers---prefer smaller $\rhoDMlocal$ than the estimates from local observations---those marked with triangles.
This could be an indication in favour of an oblate dark matter halo (see ref.~\cite{Read:2014qva} or the discussion in section~\ref{sec:spatial-coverage}).
However, a closer look at the different studies shows us that there could be important systematic uncertainties affecting the $\rhoDMlocal$ estimates, depending on the chosen tracer population, baryonic model, applied priors, and volume selection cuts.
Moreover, the results of those studies that allowed the minor-major axis ratio $q$ to vary in their analysis (e.g., \cite{Wegg_2019,Hattori:2012.03908}) are consistent with a spherical or a very-close-to-spherical dark matter halo.
Therefore, something more than just assuming a non-spherical dark matter halo is needed in order to understand the differences between the estimates shown in figure~\ref{fig:rhoDM-all-vertical}.

When comparing local and global studies, statistical uncertainties are typically larger for the former.
However, some of the global studies only considered one profile shape for each Galactic component, in particular for the baryonic components, which have occasionally been fully fixed in the analyses. 
The Galactic modelling can be made more flexible by repeating the analysis with different profile shapes for the tracer population and the respective matter components. Typically, this increases the uncertainty of inferred quantities such as $\rhoDMlocal$; this also becomes a way to quantify the systematic uncertainty associated with fixed tracer or matter density profiles. Generally, testing several profile shapes is a good practice because it allows us to reach more robust conclusions.
Furthermore, in order to appropriately interpret the quoted $\rhoDMlocal$ values from global studies, it is important to know the value of $R_\odot$ to which they apply, since a smaller $R_\odot$  results in a larger $\rhoDMlocal$ for the same Galactic model.
This is one of the reasons for the broad $2\sigma$ range of $\rhoDMlocal$ values from \cite{Benito:2019ngh}\footnote{Their recent update \cite{Benito:2020lgu} also shows similar large uncertainties with a fixed $R_\odot$ value. However, this new study widely varies the peculiar velocity of the Sun in the azimuth direction, which causes an equivalent effect to varying $R_\odot$.} in table~\ref{Tab:rhoDM-values-rot-curve}.

Differences in the estimated $\rhoDMlocal$ between studies of the same category can be due to a different choice of tracer population, a different baryonic model, or a different spatial volume.
However, the results do not seem to be significantly affected by the specific choice of method.
For example, ref.~\cite{Sivertsson:2017rkp} found dissimilar estimates from their $\alpha$-young and their $\alpha$-old samples, to which the same analysis approach was applied.
As discussed in their paper, the discrepancy could be caused by a bad modelling of the tilt term for the $\alpha$-old population, by a non-negligible effect from the otherwise neglected rotation curve term, or by disequilibrium in the stellar samples.
These possible sources of discrepancy would affect the $\alpha$-old population more than the $\alpha$-young population, since the latter has a larger vertical scale height.
Because of this larger scale height, the contribution from the rotation curve term could become relevant at the highest $|z|$. In addition, being closer to the Galactic plane, the $\alpha$-young population would probably recover equilibrium conditions faster than the $\alpha$-old stars from the passage of a possible disturber through the disc.
For all these reasons, ref.~\cite{Sivertsson:2017rkp} quoted as their preferred $\rhoDMlocal$ estimate the result of their $\alpha$-young analysis that included the tilt term.

Another source of deviation between the results of different analyses is the choice of applied priors.
Reference \cite{Guo:2020rcv} studied how $\rhoDMlocal$ depended on different modelling aspects, among which they found $\rhoDMlocal$ to be highly dependent on a prior to anchor the stellar vertical distribution. They studied the effect of switching between a prior on the total stellar surface density, $\Sigma_*$, and a prior on the stellar volume density at $z=0$, $\rho_* (z=0)$. The obtained $\rhoDMlocal$ was smaller with the latter choice, being even compatible with a value as small as $0.04\,\mathrm{GeV/cm^3}$ at $1\sigma$ if the prior on $\rho_* (z=0)$ was chosen to be flat.

The spatial volume that is analysed can also induce systematic uncertainties in the resulting $\rhoDMlocal$. 
Reference \cite{Guo:2020rcv} showed that $\rhoDMlocal$ is reasonably stable when changing the azimuthal cuts, but both refs.~\cite{Guo:2020rcv,Salomon:2020eer} found a large dependence on the chosen vertical volume, in particular when stars from either the Galactic north or south are used. 
In ref.~\cite{Salomon:2020eer} it was shown that the Galactic north seems to be more perturbed than the south, which appears to bias $\rhoDMlocal$ towards low values when observations from the Galactic south are analysed.
However, ref.~\cite{Bienayme:2014kva}, whose study was also based on Galactic south observations, preferred a larger $\rhoDMlocal$ value.

%%%%%%%%%%%%%%%%%%%%%%%%%%%%%%%%%%%%%%%%%%%%%%%%%%%%%%%%%%%%%%%%%%%%%%
\subsection{Galactic disequilibria}\label{sec:disequilibria}

The impact on the determination of $\rhoDMlocal$ caused by the perturbance of the local environment has been studied in several works. 
For instance, refs. \cite{Laporte:1808.00451,Haines:1903.00607} considered the effect of the passage of a heavy object---such as a dwarf galaxy---through the Galactic disc, and ref. \cite{Banik:2016yqm} showed that vertical breathing mode perturbations can impose systematic errors of the order of $25\%$ in the estimated $\rhoDMlocal$.

Observationally, there are clear indications that the local environment is disturbed, as it can be inferred from the vertical asymmetry in the stellar number counts \cite{Yanny:2013pqi,Bennett:1809.03507}, local vertical waves \cite{Widrow:2012wu,Gomez:2012rd,Williams:2013pbz,Carlin:2013eba,Widrow:2014jxa}, or local phase-space substructures \cite{Antoja_2018,Hunt:1806.02832,Khoperskov:1811.09205,Lopez-Corredoira:2001.05455,Necib:2018iwb,Belokurov:1802.03414}.
However, it is not yet clear to what extent the cause of local disturbances comes from the Galactic bar \cite{Hunt:1806.02832,Khoperskov:1811.09205}, spiral structures in the disc \cite{Monari:1606.06785,Michtchenko:1608.08991,Binney:1807.09819}, or the tidal interaction with the Sagittarius dwarf and past Galactic mergers \cite{Gomez:2012rd,Laporte:1808.00451,Necib:2018iwb,Vasiliev:2009.10726}. The asymmetry seen in \cite{Hinkel:2020tii}, for example, would require the effect of both the Galactic bar and the large and small Magellanic clouds.

Disequilibrium can affect different stellar populations unequally. Estimates based on different tracer populations can be biased in different ways, and certain assumptions might be valid for one population but not for another. This can be the case even if two tracer populations are currently in a steady state; for example, if their respective number density distributions are not well described by the same profile shapes.

Also the very local region, within $\lesssim 200 \,\mathrm{pc}$ from the Sun, is showing signs of disequilibrium.
References \cite{Widmark:2018ylf,Widmark:2020vqi} studied the populations of F and G type stars in the local spatial volume, using different data cuts to construct their stellar tracer population samples. Reference \cite{Widmark:2018ylf} considered the volume of a spherical shell defined by heliocentric distances between $100\,\mathrm{pc}$ and $200\,\mathrm{pc}$, constructing eight separate stellar samples of $\sim 25\,000$. Reference \cite{Widmark:2020vqi} went further in distance, reaching heights $400\,\mathrm{pc}$ above and below the Sun, and divided the Galactic plane into 40 spatially independent sub-regions.
The inferred total mass distribution in refs. \cite{Widmark:2018ylf,Widmark:2020vqi} was compared to a baryonic model (similar to e.g. \cite{McKee:2015hwa,Schutz:2017tfp,Sivertsson:2017rkp}), resulting in a significant mass surplus with a small vertical scale height of $|z| < 60\,\mathrm{pc}$.
However, the latter study \cite{Widmark:2020vqi} also showed that the inferred matter density distribution decays very quickly with height, in a way that is inconsistent with the observed scale height of the stellar disk. As such, the results of ref. \cite{Widmark:2020vqi} do no imply a mass excess per se, but a matter density distribution that is highly concentrated to the Galactic mid-plane. Because this is incompatible with the stellar disk, they concluded that their result must be biased by an incorrect assumption of a steady state. They speculate that their result could be explained by a breathing mode that is currently in it's most compressed state. Such a configuration would not be detectable by comparing the momentary motion of stars above and below the mid-plane---as the oscillation is at a turning point between compression and expansion---but would still bias this type of dynamical mass measurement.

%%%%%%%%%%%%%%%%%%%%%%%%%%%%%%%%%%%%%%%%%%%%%%%%%%%%%%%%%%%%%%%%%%%%%%
\subsection{Symmetry}\label{sec:importance-symmetry}

Non-symmetrical features and Galactic disequilibrium are connected.
Past and ongoing events in the formation history of the Milky Way that 
departs from a steady state can also manifest as some Galactic asymmetry.
For example, the passage of a heavy object through the Galactic plane distorts the phase space distribution of close stars.
Furthermore, the remnants of a past Galactic merger would currently be part of the Milky Way, possibly as a population with a distinct phase space distribution.

We know that local axisymmetry is invalid in the solar neighbourhood, as is seen for example in \cite{Bovy:1610.07610,Hinkel:2020tii}, as well as in many of the studies discussed in section \ref{sec:disequilibria}.
Similarly, the local north-south mirror symmetry seems to be broken---e.g., \cite{Guo:2020rcv,Salomon:2020eer}.

In an attempt to test the impact of non-axisymmetrical features on the estimated $\rhoDMlocal$ value,
ref.~\cite{Guo:2020rcv} divided the dwarf stars they used as tracers into different azimuthal bins, showing that the inferred $\rhoDMlocal$ was consistent if the analysis was otherwise unchanged. However, a strong variation due to an asymmetry in the vertical direction was found, in the sense that the Galactic north and south gave different results for $\rhoDMlocal$.
This result is consistent with the values found by \cite{Salomon:2020eer}.
The north-south asymmetry could be due to internal or external disturbances, such as the presence of Galactic spiral arms \cite{Monari:1606.06785}, the last passage of the Sagittarius dwarf \cite{Laporte:1808.00451} or the buckling of the Galactic bar \cite{Khoperskov:1811.09205}.

It could also be interesting to see, in future studies, if similar discrepancies to the north-south difference found by \cite{Guo:2020rcv,Salomon:2020eer} in the inferred value of $\rhoDMlocal$ are also present in the very local solar neighbourhood.

\subsection{Distribution of baryons}\label{sec:importance-baryons}

Isolating the dark matter contribution to the Galactic gravitational potential is key when inferring $\rhoDMlocal$.
Our local environment in the Milky Way is dominated by baryons, whose modelling plays a crucial role in local studies, particularly when analysing the vertical kinematics of tracer stars. Regarding the contribution to circular velocities, it is not completely clear at which specific radius dark matter starts to dominate over baryons, but the baryonic contribution is surely important at $R_\odot$ and close to the Galactic plane.
Therefore, in order to provide a robust estimate of $\rhoDMlocal$, it is necessary to have a correct characterisation of the baryonic distribution in the Galaxy, especially for our local environment.

How precise the description of baryons needs to be depends on each analysis and the method applied to infer $\rhoDMlocal$.
For example, the local studies presented in section \ref{sec:rhoDM:local-and-very-local:local} need a good description of the baryonic distribution at further distances from the Galactic plane, but are less dependent on the precise shape of the baryonic matter distribution close to the mid-plane.
However, for the very local studies discussed in section \ref{sec:rhoDM:local-and-very-local:very-local}, it becomes crucial to appropriately model the baryonic distribution within $200\,\mathrm{pc}$ from the position of the Sun.
In this regard, most of the papers discussed in section \ref{sec:rhoDM:local-and-very-local:very-local} follow the same baryonic model \cite{McKee:2015hwa}, which is based on different pre-Gaia studies \cite{Flynn:2006tm,Kramer:2016dew}. This model is included either directly as part of the dynamical modelling \cite{Schutz:2017tfp,Buch:2018qdr}, or as a post-inference comparison with the results \cite{Widmark:2018ylf,Widmark:2020vqi}. However, this baryonic model suffers from some shortcomings and potentially substantial systematic errors. The model consists of twelve separate components of different types of gas (cold and warm atomic gas, molecular gas, hot ionised gas), stars in different brightness ranges, white dwarfs, and brown dwarfs. 
Each of the twelve components is assumed to be isothermal, i.e. in equilibrium and with a Gaussian vertical velocity distribution; its matter density at any height is determined by its mid-plane matter density, its vertical velocity dispersion, and the gravitational potential. To begin with, the different stellar populations' vertical velocity distributions are not well described by single Gaussians, and their number density as a function of height is not very well predicted by the baryonic model---see for example appendix A in ref.~\cite{Widmark:2020vqi}. Potentially, even more significant systematic uncertainties affect the gaseous components. Measurements of the cold gas depend on some poorly constrained quantities: the CO-to-H$_2$ conversion factor is used when measuring molecular gas, and corrections for optical depth affect the $21\, \mathrm{cm}$ based measurements of atomic gas \cite{Hessman:2015nua}. Furthermore, the spatial structure of cold gas is highly non-uniform---for example, the very local volume has a very low density of gas and dust \cite{Lallement:2003_a}. For these reasons, it is plausible that the systematic error associated with the baryonic modelling can be significantly larger than the statistical uncertainty in very local studies of $\rhoDMlocal$.

Estimates of $\rhoDMlocal$ from Galactic circular velocities are equally dependent on the baryonic description.
Given a Galactic mass model, the contribution to $v_{\rm c}(R_\odot)$ from its dark matter halo is similar to the contribution from its baryons. Therefore, in spite of the current good precision of $v_{\rm c,obs}(R)$ between $4\,\mathrm{kpc}\lesssim R \lesssim 25\,\mathrm{kpc}$ \cite{Eilers:1810.09466,Mroz:1810.02131}, a precise estimate of $\rhoDMlocal$ can not be achieved without a similarly precise understanding of the baryonic distribution.
In fact, varying the profile shape of the distribution of baryons affects more strongly the fitted value of $\rhoDMlocal$ than changing the shape of the dark matter halo. 
This is the case because of two main reasons. First, the local distribution of the different baryonic components (gas, stars and dust) still present large uncertainties, and different shapes can fit the observations. Second, it is expected that the dark matter halo does not deviate much from a spherical shape, and its contribution to $v_{\rm c}(R)$ is more robust than the contribution from baryons when varying the profile shape, at least in the relevant range of $R \sim \text{4--25}\,\mathrm{kpc}$.
In this vein, it is important to highlight that the circular velocity curve is very unlikely going to help us constrain the shape of the dark matter halo inside $\sim 4 \,\mathrm{kpc}$. This is due to the fact that circular velocity estimates so close to the Galactic centre can suffer from large systematic uncertainties, especially due to the broken axisymmetric at those radial distances \cite{Chemin:1504.01507}.

References \cite{Huang:1604.01216} and \cite{Lin:2019yux} constitute another example of the importance of baryons. 
The difference between the results of these two references are one of the largest discrepancies found in table \ref{Tab:rhoDM-values-rot-curve}.
This is somewhat surprising because the two studies have used the same data for their analyses. 
Part of this difference is due to the choice of $R_\odot$, which is smaller in ref. \cite{Lin:2019yux}. However, if we consider the best-fit dark matter halo from ref. \cite{Huang:1604.01216} and compute the dark matter density at $R_\odot = 8.0\,\mathrm{kpc}$---the same value assumed in ref.~\cite{Lin:2019yux}---the corresponding value of $\rhoDMlocal$ increases by only $0.023\,\mathrm{GeV/cm^3}$, not enough to make both results compatible at their quoted $1\sigma$ uncertainty.
Ultimately, most of the difference comes from the different choice of the Galactic mass model, in particular from the treatment of the baryonic components. 
Other works from table \ref{Tab:rhoDM-values-rot-curve} also show the impact of baryons on the results. For example, in refs.~\cite{Pato:2015dua,deSalas:2019pee} is seen that, when different baryonic models---i.e. different baryonic shapes---are fitted to the same data, the resulting $\rhoDMlocal$ usually shows a larger variance than the corresponding uncertainty of each individual analysis.

Here it is also worth mentioning the work of ref.~\cite{Salucci:2010qr}, in which most of previously discussed problems were already considered from the perspective of a $\rhoDMlocal$ inference from the circular velocity curve. 
Using common assumptions like those of the Ideal Galaxy (see section~\ref{sec:Ideal-Galaxy}) and a baryonic description dominated by a radially-exponential Galactic disc, they provided a theoretical equation (their eq.~11) for $\rhoDMlocal$, which depends on relevant parameters such as the circular velocity at the Sun's location.
We have tested this equation with the two baryonic models of ref.~\cite{deSalas:2019pee} (tagged B1 and B2, see section~\ref{sec:rhoDM:rot-curve} or the description in the original paper), finding a good agreement when the baryonic model B2, which includes a single double-exponential Galactic disc, is used. However, the results are more difficult to compare with the analysis of the baryonic model B1, which considers two Galactic discs that do not follow an exponential density.
For future circular velocity studies aiming at going beyond the approximations of the Ideal Galaxy, the equation from ref.~\cite{Salucci:2010qr} could offer a comparison with the expected $\rhoDMlocal$ value in an idealised Milky Way. 

%%%%%%%%%%%%%%%%%%%%%%%%%%%%%%%%%%%%%%%%%%%%%%%%%%%%%%%%%%%%%%%%%%%%%%
\subsection{Unconventional dark matter components}\label{sec:new-physics}

A scenario beyond the standard picture of a cold dark matter halo can affect the estimates of $\rhoDMlocal$.
One example is the possibility of the Galaxy containing a dark matter disc, that could come into existence either by dynamical friction---if the baryonic disc was formed very early, triggering the accretion of subhalos into the Galactic plane \cite{Lake:1989.AJ.98.1554,Read:2008fh}---or by a dark matter subcomponent with strong dissipative self-interactions
\cite{Fan:2013tia,Fan:2013yva}. 
The mass of an hypothetical dark matter disc can only be a small fraction of the mass of the baryonic disc \cite{Schutz:2017tfp,Buch:2018qdr}, so the dark disc contribution to the Galactic circular velocity would be very small.
Therefore, a dark disc would affect especially the local vertical kinematics studies, rather than global studies covering larger spatial volumes.
In this sense, a dark disc would create a discrepancy between global and local estimates of $\rhoDMlocal$, similar to an oblate dark matter halo.

At present, all local studies are compatible with the non-existence of a dark disc.
Even the large overdensity that was noticed in \cite{Widmark:2018ylf} is very likely due to disequilibrium effects such as a breathing mode of the Galactic disc \cite{Widmark:2020vqi}. 
Furthermore, a couple of recent studies \cite{Schutz:2017tfp,Buch:2018qdr} constrained the presence of a dark disc within a similar spatial region as ref.~\cite{Widmark:2018ylf}, severely restricting the parameter space of such an unconventional component.

Another possible contribution to $\rhoDMlocal$ is a dark matter flux associated to a passing substructure through our local region. For example, in ref.~\cite{OHare:2018trr} they considered the effect to direct detection experiments from a small contribution to $\rhoDMlocal$ from the S1 stream \cite{Myeong:2017skt}.

%%%%%%%%%%%%%%%%%%%%%%%%%%%%%%%%%%%%%%%%%%%%%%%%%%%%%%%%%%%%%%%%%%%%%%
%%%%%%%%%%%%%%%%%%%%%%%%%%%%%%%%%%%%%%%%%%%%%%%%%%%%%%%%%%%%%%%%%%%%%%
\section{Future estimates: new ideas and final Gaia release}\label{sec:new-ideas}

The analysis of the first two Gaia data releases has shown that the Milky Way is not the ideal configuration of stars that we often approximate it to be (see section \ref{sec:breaking-of-Ideal-Galaxy}). 
Hence, a precise determination of the Galactic properties, including $\rhoDMlocal$, not only requires precise data and large statistics, but also improved techniques to analyse the wealth of present and future observations.

Gaia DR2 contains more than one billion 5-dimensional astrometric observations (i.e. with measured parallax, angular position and proper motions), of which about seven million also include radial velocities \cite{Brown:2018dum}. 
These observations have been recently superseded by Gaia EDR3 \cite{Brown:2012.01533}.
Future Gaia data releases will increase the quantity of 6-dimensional observations, as well as the precision of those already included in Gaia EDR3. 
The increasing amount of statistics will help us characterise the current state of the Milky Way, as well as understand the past history that led the Galaxy to its present configuration. 
However, a reduction of statistical uncertainties does not guarantee a reduction of total uncertainties, which could instead be dominated by systematic uncertainties associated with the analysis method and modelling assumptions.

One possible tool for diagnosing and quantifying modelling uncertainties is to have a more complete description of the Galactic objects. For example, adding stellar properties like metallicity to their astrometric information enhances the ability to distinguish between stellar type and age, which can be useful to build a more sophisticated Galactic model. 

Another way of increasing the complexity and accuracy of a Galactic model is to include information about stellar accelerations. Our current knowledge of the Milky Way is mostly based on an incomplete picture of its present configuration---i.e. from current positions and velocities---with little to no information on the acceleration of its orbiting objects. 
This information would increase our knowledge of the gravitational potential, breaking its degeneracy with the time-dependent term in the Boltzmann equation.
In this direction, Gaia EDR3 observations have already been able to constrain the solar system acceleration \cite{Klioner:2012.02036}, which ref.~\cite{Bovy:2012.02169} used to obtain Galactic properties that control the dynamics of our local environment.
In addition, refs.~\cite{Ravi:2018vqd,Silverwood:2018qra} proposed the use of Doppler spectroscopy to observe the time derivative of stellar radial velocities, which would help us constrain their Galactic accelerations.
However, the required precision is beyond the capability of current and planned spectrographs, which would have to distinguish the Galactic acceleration from contaminants caused by stellar or planetary companions and noise sources such as stellar activity \cite{Ravi:2018vqd}.

Another novel approach to measure the matter density of the Galactic disc---and therefore also $\rhoDMlocal$---involves using stellar streams, as proposed in ref. \cite{Widmark:2020zad}.
A stellar stream is an elongated structure formed from the tidal disruption of a satellite, such as a globular cluster or a dwarf galaxy orbiting the Milky Way. A stellar stream is formed from stars stripped from its progenitor, typically with small variations in the stars' total energy and angular momentum, such that they are all roughly on the same Galactic orbit. For a stellar stream passing through or close to the Galactic disc, the gravitational potential and mass of the disc can be inferred by fitting this orbit. Such a measurement is especially useful because its complementary: it does not rely on the same assumption of a steady state for the stellar disc, which methods using disc tracer stars rely upon. The ideal stellar stream for the method presented in ref. \cite{Widmark:2020zad} would preferably be small and dynamically cold, extend at least a few hundred parsecs in length, and be located at a distance of at most a few kiloparsecs from the Sun. Such a stream has not yet been observed, but this could change soon, especially considering that most current stream finding algorithms mask the Galactic plane for reasons of computational tractability \cite{Malhan:2018_streamfinder,Borsato:1907.02527}.

As we have mentioned in the introduction, one of the reasons why a precise estimate of $\rhoDMlocal$ is very important is because it constitutes a relevant quantity for direct detection searches of dark matter. 
However, the value of $\rhoDMlocal$ is degenerate with the cross section $\sigma$ of the dark matter-nucleus interaction---see eq.~\eqref{eq:differential-recoil-rate-SI}---hence the need of a precise $\rhoDMlocal$ in order to make a correct interpretation of a positive signal.
As shown in \cite{Kavanagh:2020cvn}, this degeneracy can be broken provided that a) the scattering of dark matter particles in the detector is frequent enough, and b) the detector is able to observe the diurnal variation of the dark matter flux when passing through the Earth. Under these hypotheses, it would be possible to obtain a simultaneous estimate of both quantities: $\rhoDMlocal$ and $\sigma$.
This scenario is discussed in \cite{Kavanagh:2020cvn}, where they mention that $\rhoDMlocal$ could be measured with a $20\%$ uncertainty for a sub-GeV dark matter-proton spin-independent cross section of $\sim 10^{-32}\,\mathrm{cm^2}$.

Finally, in the speculative regime, the Vogage 2050 white paper \cite{Berge:2019zjj} suggests probing the gravitational potential by tracking a spacecraft equipped with an accelerometer and an atomic clock, and launched out of the gravitational reach of the Solar System. 
However, the success of this idea hinges on a significant technological improvement, such as the development of innovative propulsion techniques.

%%%%%%%%%%%%%%%%%%%%%%%%%%%%%%%%%%%%%%%%%%%%%%%%%%%%%%%%%%%%%%%%%%%%%%
%%%%%%%%%%%%%%%%%%%%%%%%%%%%%%%%%%%%%%%%%%%%%%%%%%%%%%%%%%%%%%%%%%%%%%
\section{Conclusions}\label{sec:conclusions}

In the last few years, there have been an increasing number of publications related to $\rhoDMlocal$. This increase has been partly triggered by the observations of the Gaia satellite mission, in particular from its second data release \cite{Brown:2018dum}. Gaia's observations have allowed us to improve our knowledge of the Milky Way, unveiling the complexity of our Galaxy. 

In this report, we have covered the result of recent $\rhoDMlocal$ estimates, which we summarised in figure \ref{fig:rhoDM-all-vertical}. The value of $\rhoDMlocal$ has been estimated from a large variety of methods, using different data and techniques, which makes these studies largely complementary to each other. The results of local analyses (section \ref{sec:rhoDM:local-and-very-local}) prefer a range of $\rhoDMlocal \simeq \text{0.4--0.6}\,\mathrm{GeV/cm^3}$, which is slightly higher than the preferred range of $\rhoDMlocal \simeq \text{0.3--0.5}\,\mathrm{GeV/cm^3}$ from most global analyses (sections \ref{sec:rhoDM:rot-curve} and \ref{sec:rhoDM:DF-global}). 
Overall, there is a good agreement between the different studies, although some tensions remain. There can be many reasons for these discrepancies, arising from systematic uncertainties associated with the dynamical modelling and its underlying assumptions---including the distribution of baryons---or from the data itself (section~\ref{sec:breaking-of-Ideal-Galaxy}).
Many studies have benefited from Gaia data, but we are still relying on the canonical assumptions of a steady state, axisymmetry and mirror symmetry across the Galactic plane (see section \ref{sec:Ideal-Galaxy}), although the latter has been relaxed in a few recent studies.

With the Gaia mission, the data available for analysis has surpassed our current dynamical modelling techniques; for example, the data set shows clear signatures of time-varying dynamical structures, breaking the assumption of a steady state which practically all estimates of $\rhoDMlocal$ rely upon. As such, we have probably reached a point were our dynamical models are dominated by systematic, rather than statistical, uncertainties. For this reason, robust estimates of $\rhoDMlocal$ require more sophisticated modelling techniques where these assumptions are relaxed, or, at the very least, their combined biases are carefully diagnosed and quantified.
The community is looking forward to the final Gaia data release and all the interesting discoveries and important information that will be revealed about the Milky Way. Hopefully, our modelling efforts will step to the challenge of this vast and complex data set, such that the next review on the subject of $\rhoDMlocal$ can report on a deeper and more accurate knowledge of our Galaxy.

%%%%%%%%%%%%%%%%%%%%%%%%%%%%%%%%%%%%%%%%%%%%%%%%%%%%%%%%%%%%%%%%%%%%%%
%%%%%%%%%%%%%%%%%%%%%%%%%%%%%%%%%%%%%%%%%%%%%%%%%%%%%%%%%%%%%%%%%%%%%%

\acknowledgments
Pablo F. de Salas acknowledge support by the Vetenskapsr{\aa}det (Swedish Research Council) through contract No. 638-2013-8993 and the Oskar Klein Centre for Cosmoparticle Physics. Axel Widmark acknowledges support from the Carlsberg Foundation via a Semper Ardens grant (CF15-0384).

\bibliographystyle{utcaps}
%\bibliography{refs}

\begin{thebibliography}{100}

\bibitem{Bertone:2016nfn}
G.~Bertone and D.~Hooper, ``{History of dark matter},''
  \href{http://dx.doi.org/10.1103/RevModPhys.90.045002}{{\em Rev. Mod. Phys.}
  {\bfseries 90} (2018)  045002},
  \href{http://arxiv.org/abs/1605.04909}{{\ttfamily arXiv:1605.04909
  [astro-ph.CO]}}.

\bibitem{Aghanim:2018eyx}
{\bfseries Planck} Collaboration, N.~Aghanim {\em et al.}, ``{Planck 2018
  results. VI. Cosmological parameters},''
  \href{http://dx.doi.org/10.1051/0004-6361/201833910}{{\em Astron. Astrophys.}
  {\bfseries 641} (2020)  A6},
  \href{http://arxiv.org/abs/1807.06209}{{\ttfamily arXiv:1807.06209
  [astro-ph.CO]}}.

\bibitem{Bergstrom:2000pn}
L.~Bergstr{\"o}m, ``{Nonbaryonic dark matter: Observational evidence and
  detection methods},''
  \href{http://dx.doi.org/10.1088/0034-4885/63/5/2r3}{{\em Rept. Prog. Phys.}
  {\bfseries 63} (2000)  793},
  \href{http://arxiv.org/abs/hep-ph/0002126}{{\ttfamily arXiv:hep-ph/0002126
  [hep-ph]}}.

\bibitem{Bertone:2004pz}
G.~Bertone, D.~Hooper, and J.~Silk, ``{Particle dark matter: Evidence,
  candidates and constraints},''
  \href{http://dx.doi.org/10.1016/j.physrep.2004.08.031}{{\em Phys. Rept.}
  {\bfseries 405} (2005)  279--390},
  \href{http://arxiv.org/abs/hep-ph/0404175}{{\ttfamily arXiv:hep-ph/0404175
  [hep-ph]}}.

\bibitem{Roszkowski:2017nbc}
L.~Roszkowski, E.~M. Sessolo, and S.~Trojanowski, ``{WIMP dark matter
  candidates and searches---current status and future prospects},''
  \href{http://dx.doi.org/10.1088/1361-6633/aab913}{{\em Rept. Prog. Phys.}
  {\bfseries 81} (2018)  066201},
  \href{http://arxiv.org/abs/1707.06277}{{\ttfamily arXiv:1707.06277
  [hep-ph]}}.

\bibitem{Goodman:2010ku}
J.~Goodman, M.~Ibe, A.~Rajaraman, W.~Shepherd, T.~M. Tait, and H.-B. Yu,
  ``{Constraints on Dark Matter from Colliders},''
  \href{http://dx.doi.org/10.1103/PhysRevD.82.116010}{{\em Phys. Rev. D}
  {\bfseries 82} (2010)  116010},
  \href{http://arxiv.org/abs/1008.1783}{{\ttfamily arXiv:1008.1783 [hep-ph]}}.

\bibitem{Klasen:2015uma}
M.~Klasen, M.~Pohl, and G.~Sigl, ``{Indirect and direct search for dark
  matter},'' \href{http://dx.doi.org/10.1016/j.ppnp.2015.07.001}{{\em Prog.
  Part. Nucl. Phys.} {\bfseries 85} (2015)  1--32},
  \href{http://arxiv.org/abs/1507.03800}{{\ttfamily arXiv:1507.03800
  [hep-ph]}}.

\bibitem{Undagoitia:2015gya}
T.~Marrod{\'a}n~Undagoitia and L.~Rauch, ``{Dark matter direct-detection
  experiments},'' \href{http://dx.doi.org/10.1088/0954-3899/43/1/013001}{{\em
  J. Phys. G} {\bfseries 43} (2016)  013001},
  \href{http://arxiv.org/abs/1509.08767}{{\ttfamily arXiv:1509.08767
  [physics.ins-det]}}.

\bibitem{Green:2011bv}
A.~M. Green, ``{Astrophysical uncertainties on direct detection experiments},''
  \href{http://dx.doi.org/10.1142/S0217732312300042}{{\em Mod. Phys. Lett. A}
  {\bfseries 27} (2012)  1230004},
  \href{http://arxiv.org/abs/1112.0524}{{\ttfamily arXiv:1112.0524
  [astro-ph.CO]}}.

\bibitem{Necib:2018iwb}
L.~Necib, M.~Lisanti, and V.~Belokurov, ``{Inferred Evidence For Dark Matter
  Kinematic Substructure with SDSS-Gaia},''
  \href{http://dx.doi.org/10.3847/1538-4357/ab095b}{{\em {Astrophys. J.}}
  {\bfseries {874}} (2019)  {3}},
  \href{http://arxiv.org/abs/1807.02519}{{\ttfamily arXiv:1807.02519
  [astro-ph.GA]}}.

\bibitem{Necib:2018igl}
L.~Necib {\em et al.}, ``{Under the Firelight: Stellar Tracers of the Local
  Dark Matter Velocity Distribution in the Milky Way},''
  \href{http://dx.doi.org/10.3847/1538-4357/ab3afc}{{\em {Astrophys. J.}}
  {\bfseries {883}} (2019)  {27}},
  \href{http://arxiv.org/abs/1810.12301}{{\ttfamily arXiv:1810.12301
  [astro-ph.GA]}}.

\bibitem{Bozorgnia:2018pfa}
N.~Bozorgnia {\em et al.}, ``{On the correlation between the local dark matter
  and stellar velocities},''
  \href{http://dx.doi.org/10.1088/1475-7516/2019/06/045}{{\em \jcap} {\bfseries
  06} (2019)  045}, \href{http://arxiv.org/abs/1811.11763}{{\ttfamily
  arXiv:1811.11763 [astro-ph.GA]}}.

\bibitem{Read:2014qva}
J.~I. Read, ``{The Local Dark Matter Density},''
  \href{http://dx.doi.org/10.1088/0954-3899/41/6/063101}{{\em J. Phys. G}
  {\bfseries 41} (2014)  063101},
  \href{http://arxiv.org/abs/1404.1938}{{\ttfamily arXiv:1404.1938
  [astro-ph.GA]}}.

\bibitem{Bland-Hawthorn:1602.07702}
J.~Bland-Hawthorn and O.~Gerhard, ``{The Galaxy in Context: Structural,
  Kinematic and Integrated Properties},''
  \href{http://dx.doi.org/10.1146/annurev-astro-081915-023441}{{\em {Ann. Rev.
  Astron. Astrophys.}} {\bfseries 54} (2016)  529},
  \href{http://arxiv.org/abs/1602.07702}{{\ttfamily arXiv:1602.07702
  [astro-ph.GA]}}.

\bibitem{Read:2008fh}
J.~I. Read, G.~Lake, O.~Agertz, and V.~P. Debattista, ``{Thin, thick and dark
  discs in LCDM},''
  \href{http://dx.doi.org/10.1111/j.1365-2966.2008.13643.x}{{\em \mnras}
  {\bfseries 389} (2008)  1041},
  \href{http://arxiv.org/abs/0803.2714}{{\ttfamily arXiv:0803.2714
  [astro-ph]}}.

\bibitem{Purcell:2009yp}
C.~W. Purcell, J.~S. Bullock, and M.~Kaplinghat, ``{The Dark Disk of the Milky
  Way},'' \href{http://dx.doi.org/10.1088/0004-637X/703/2/2275}{{\em Astrophys.
  J.} {\bfseries 703} (2009)  2275},
  \href{http://arxiv.org/abs/0906.5348}{{\ttfamily arXiv:0906.5348
  [astro-ph.GA]}}.

\bibitem{Fan:2013tia}
J.~Fan, A.~Katz, L.~Randall, and M.~Reece, ``{Dark-Disk Universe},''
  \href{http://dx.doi.org/10.1103/PhysRevLett.110.211302}{{\em Phys. Rev.
  Lett.} {\bfseries 110} (2013)  211302},
  \href{http://arxiv.org/abs/1303.3271}{{\ttfamily arXiv:1303.3271 [hep-ph]}}.

\bibitem{Fan:2013yva}
J.~Fan, A.~Katz, L.~Randall, and M.~Reece, ``{Double-Disk Dark Matter},''
  \href{http://dx.doi.org/10.1016/j.dark.2013.07.001}{{\em Phys. Dark Univ.}
  {\bfseries 2} (2013)  139}, \href{http://arxiv.org/abs/1303.1521}{{\ttfamily
  arXiv:1303.1521 [astro-ph.CO]}}.

\bibitem{Kapteyn:1922zz}
J.~C. Kapteyn, ``First Attempt at a Theory of the Arrangement and Motion of the
  Sidereal System,'' \href{http://dx.doi.org/10.1086/142670}{{\em Astrophys.
  J.} {\bfseries 55} (1922)  302}.

\bibitem{Jeans_1922}
J.~H. Jeans, ``{The Motions of Stars in a Kapteyn-Universe},''
  \href{http://dx.doi.org/10.1093/mnras/82.3.122}{{\em \mnras} {\bfseries 82}
  (1922)  122}.

\bibitem{BinneyTremaine:book}
J.~{Binney} and S.~{Tremaine}, {\em {Galactic Dynamics: Second Edition}}.
\newblock Princeton University Press, 2008.

\bibitem{Belokurov:1802.03414}
V.~{Belokurov}, D.~{Erkal}, N.~W. {Evans}, S.~E. {Koposov}, and A.~J. {Deason},
  ``{Co-formation of the disc and the stellar halo},''
  \href{http://dx.doi.org/10.1093/mnras/sty982}{{\em \mnras} {\bfseries 478}
  (2018)  611--619}, \href{http://arxiv.org/abs/1802.03414}{{\ttfamily
  arXiv:1802.03414 [astro-ph.GA]}}.

\bibitem{Helmi:1806.06038}
A.~{Helmi}, C.~{Babusiaux}, H.~H. {Koppelman}, D.~{Massari}, J.~{Veljanoski},
  and A.~G.~A. {Brown}, ``{The merger that led to the formation of the Milky
  Way's inner stellar halo and thick disk},''
  \href{http://dx.doi.org/10.1038/s41586-018-0625-x}{{\em Nature} {\bfseries
  563} (2018) no.~7729, 85--88},
  \href{http://arxiv.org/abs/1806.06038}{{\ttfamily arXiv:1806.06038
  [astro-ph.GA]}}.

\bibitem{Antoja_2018}
T.~Antoja, A.~Helmi, M.~Romero-G{\'{o}}mez, D.~Katz, {\em et al.}, ``{A
  dynamically young and perturbed Milky Way disk},''
  \href{http://dx.doi.org/10.1038/s41586-018-0510-7}{{\em Nature} {\bfseries
  561} (2018) no.~7723, 360}.

\bibitem{Khoperskov:1811.09205}
S.~Khoperskov, P.~D. Matteo, O.~Gerhard, {\em et al.}, ``{The echo of the bar
  buckling: phase-space spirals in Gaia Data Release 2},''
  \href{http://dx.doi.org/10.1051/0004-6361/201834707}{{\em Astron. Astrophys.}
  {\bfseries 622} (2019)  L6},
  \href{http://arxiv.org/abs/1811.09205}{{\ttfamily arXiv:1811.09205
  [astro-ph.GA]}}.

\bibitem{Binney:1807.09819}
J.~{Binney} and R.~{Sch{\"o}nrich}, ``{The origin of the Gaia phase-plane
  spiral},'' \href{http://dx.doi.org/10.1093/mnras/sty2378}{{\em MNRAS}
  {\bfseries 481} (2018)  1501--1506},
  \href{http://arxiv.org/abs/1807.09819}{{\ttfamily arXiv:1807.09819
  [astro-ph.GA]}}.

\bibitem{Guo:2020rcv}
R.~Guo, C.~Liu, S.~Mao, X.-X. Xue, R.~Long, and L.~Zhang, ``{Measuring the
  local dark matter density with LAMOST DR5 and Gaia DR2},''
  \href{http://dx.doi.org/10.1093/mnras/staa1483}{{\em \mnras} {\bfseries 495}
  (2020)  4828--4844}, \href{http://arxiv.org/abs/2005.12018}{{\ttfamily
  arXiv:2005.12018 [astro-ph.GA]}}.

\bibitem{Salomon:2020eer}
J.-B. Salomon, O.~Bienaym{\'e}, C.~Reyl{\'e}, A.~C. Robin, and B.~Famaey,
  ``{Kinematics and dynamics of Gaia red clump stars},''
  \href{http://dx.doi.org/10.1051/0004-6361/202038535}{{\em Astron. Astrophys.}
  {\bfseries 643} (2020)  A75},
  \href{http://arxiv.org/abs/2009.04495}{{\ttfamily arXiv:2009.04495
  [astro-ph.GA]}}.

\bibitem{Bovy:2014vfa}
J.~Bovy, ``{galpy: A Python Library for Galactic Dynamics},''
  \href{http://dx.doi.org/10.1088/0067-0049/216/2/29}{{\em Astrophys. J.
  Suppl.} {\bfseries 216} (2015)  29},
  \href{http://arxiv.org/abs/1412.3451}{{\ttfamily arXiv:1412.3451
  [astro-ph.GA]}}.

\bibitem{McMillan:1608.00971}
P.~J. {McMillan}, ``{The mass distribution and gravitational potential of the
  Milky Way},'' \href{http://dx.doi.org/10.1093/mnras/stw2759}{{\em \mnras}
  {\bfseries 465} (2017)  76--94},
  \href{http://arxiv.org/abs/1608.00971}{{\ttfamily arXiv:1608.00971
  [astro-ph.GA]}}.

\bibitem{Syer:1996uv}
D.~Syer and S.~Tremaine, ``{Made to measure N body systems},''
  \href{http://dx.doi.org/10.1093/mnras/282.1.223}{{\em \mnras} {\bfseries 282}
  (1996)  223}, \href{http://arxiv.org/abs/astro-ph/9605061}{{\ttfamily
  arXiv:astro-ph/9605061 [astro-ph]}}.

\bibitem{Bovy:1704.03884}
J.~{Bovy}, D.~{Kawata}, and J.~A.~S. {Hunt}, ``{Made-to-measure modelling of
  observed galaxy dynamics},''
  \href{http://dx.doi.org/10.1093/mnras/stx2402}{{\em \mnras} {\bfseries 473}
  (2018)  2288--2303}, \href{http://arxiv.org/abs/1704.03884}{{\ttfamily
  arXiv:1704.03884 [astro-ph.GA]}}.

\bibitem{Nitschai:1909.05269}
M.~S. {Nitschai}, M.~{Cappellari}, and N.~{Neumayer}, ``{First Gaia dynamical
  model of the Milky Way disc with six phase space coordinates: a test for
  galaxy dynamics},'' \href{http://dx.doi.org/10.1093/mnras/staa1128}{{\em
  \mnras} {\bfseries 494} (2020)  6001--6011},
  \href{http://arxiv.org/abs/1909.05269}{{\ttfamily arXiv:1909.05269
  [astro-ph.GA]}}.

\bibitem{Wegg_2019}
C.~Wegg, O.~Gerhard, and M.~Bieth, ``{The gravitational force field of the
  Galaxy measured from the kinematics of {RR} Lyrae in Gaia},''
  \href{http://dx.doi.org/10.1093/mnras/stz572}{{\em \mnras} {\bfseries 485}
  (2019)  3296--3316}, \href{http://arxiv.org/abs/1806.09635}{{\ttfamily
  arXiv:1806.09635 [astro-ph.GA]}}.

\bibitem{Hattori:2012.03908}
K.~Hattori, M.~Valluri, and E.~Vasiliev, ``{Action-based distribution function
  modelling for constraining the shape of the Galactic dark matter halo},''
  \href{http://arxiv.org/abs/2012.03908}{{\ttfamily arXiv:2012.03908
  [astro-ph.GA]}}.

\bibitem{Dehnen:1999ea}
W.~Dehnen, ``{Simple distribution functions for stellar disks},''
  \href{http://dx.doi.org/10.1086/301010}{{\em Astron. J.} {\bfseries 118}
  (1999)  1201}.

\bibitem{Binney:2011xa}
J.~Binney and P.~McMillan, ``{Models of our Galaxy II},''
  \href{http://dx.doi.org/10.1111/j.1365-2966.2011.18268.x}{{\em \mnras}
  {\bfseries 413} (2011)  1889},
  \href{http://arxiv.org/abs/1101.0747}{{\ttfamily arXiv:1101.0747
  [astro-ph.GA]}}.

\bibitem{Vasiliev:1802.08239-agama}
E.~Vasiliev, ``{AGAMA: Action-based galaxy modelling architecture},''
  \href{http://dx.doi.org/10.1093/mnras/sty2672}{{\em \mnras} {\bfseries 482}
  (2018)  1525}, \href{http://arxiv.org/abs/1802.08239}{{\ttfamily
  arXiv:1802.08239 [astro-ph.GA]}}.

\bibitem{Binney:2014jda}
J.~Binney, ``{Self-consistent flattened isochrone models},''
  \href{http://dx.doi.org/10.1093/mnras/stu297}{{\em \mnras} {\bfseries 440}
  (2014)  787--798}, \href{http://arxiv.org/abs/1402.2512}{{\ttfamily
  arXiv:1402.2512 [astro-ph.GA]}}.

\bibitem{Posti:1411.7897}
L.~{Posti}, J.~{Binney}, C.~{Nipoti}, and L.~{Ciotti}, ``{Action-based
  distribution functions for spheroidal galaxy components},''
  \href{http://dx.doi.org/10.1093/mnras/stu2608}{{\em \mnras} {\bfseries 447}
  (2015)  3060--3068}, \href{http://arxiv.org/abs/1411.7897}{{\ttfamily
  arXiv:1411.7897 [astro-ph.GA]}}.

\bibitem{Cuddeford:1991MNRAS.253..414C}
P.~{Cuddeford}, ``{An analytic inversion for anisotropic spherical galaxies},''
  \href{http://dx.doi.org/10.1093/mnras/253.3.414}{{\em \mnras} {\bfseries 253}
  (1991)  414--426}.

\bibitem{Binney:2013mhf}
J.~Binney, ``{Dynamics for Galactic Archaeology},''
  \href{http://dx.doi.org/10.1016/j.newar.2013.08.001}{{\em New Astron.Rev.}
  {\bfseries 57} (2013)  29--51},
  \href{http://arxiv.org/abs/1309.2794}{{\ttfamily arXiv:1309.2794
  [astro-ph.GA]}}.

\bibitem{Sanders:1511.08213}
J.~L. Sanders and J.~Binney, ``{A review of action estimation methods for
  galactic dynamics},'' \href{http://dx.doi.org/10.1093/mnras/stw106}{{\em
  MNRAS} {\bfseries 457} (2016)  2107},
  \href{http://arxiv.org/abs/1511.08213}{{\ttfamily arXiv:1511.08213
  [astro-ph.GA]}}.

\bibitem{Lundmark:1930}
K.~{Lundmark}, ``{{\"U}ber die Bestimmung der Entfernungen, Dimensionen, Massen
  und Dichtigkeit fur die n{\"a}chstgelegenen anagalacktischen
  Sternsysteme.},'' {\em Meddelanden fran Lunds Astronomiska Observatorium
  Serie I} {\bfseries 125} (1930)  1--13.

\bibitem{Rubin:1980zd}
V.~C. Rubin, N.~Thonnard, and W.~K. Ford, Jr., ``{Rotational properties of 21
  SC galaxies with a large range of luminosities and radii, from NGC 4605 /R =
  4kpc/ to UGC 2885 /R = 122 kpc/},''
\href{http://dx.doi.org/10.1086/158003}{{\em Astrophys. J.} {\bfseries 238}
  (1980)  471}.
%%CITATION = ASJOA,238,471;%%.

\bibitem{Sofue:2020rnl}
Y.~Sofue, ``{Rotation Curve of the Milky Way and the Dark Matter Density},''
  \href{http://dx.doi.org/10.3390/galaxies8020037}{{\em Galaxies} {\bfseries 8}
  (2020)  37}, \href{http://arxiv.org/abs/2004.11688}{{\ttfamily
  arXiv:2004.11688 [astro-ph.GA]}}.

\bibitem{Eilers:1810.09466}
A.-C. {Eilers}, D.~W. {Hogg}, H.-W. {Rix}, and M.~K. {Ness}, ``{The Circular
  Velocity Curve of the Milky Way from 5 to 25 kpc},''
  \href{http://dx.doi.org/10.3847/1538-4357/aaf648}{{\em Astrophys. J.}
  {\bfseries 871} (2019)  120},
  \href{http://arxiv.org/abs/1810.09466}{{\ttfamily arXiv:1810.09466
  [astro-ph.GA]}}.

\bibitem{BinneyMerrifield:book}
J.~{Binney} and M.~{Merrifield}, {\em {Galactic Astronomy}}.
\newblock Princeton University Press, 1998.

\bibitem{Brown:2012.01533}
{\bfseries Gaia} Collaboration, A.~G.~A. Brown {\em et al.}, ``{Gaia Early Data
  Release 3: Summary of the contents and survey properties},''
  \href{http://arxiv.org/abs/2012.01533}{{\ttfamily arXiv:2012.01533
  [astro-ph.GA]}}.

\bibitem{Fabricius:2012.06242}
C.~Fabricius {\em et al.}, ``{Gaia Early Data Release 3 -- Catalogue
  validation},'' \href{http://arxiv.org/abs/2012.06242}{{\ttfamily
  arXiv:2012.06242 [astro-ph.GA]}}.

\bibitem{Oort:1932_a}
J.~H. {Oort}, ``{The force exerted by the stellar system in the direction
  perpendicular to the galactic plane and some related problems},'' {\em
  Bulletin of the Astronomical Institutes of the Netherlands} {\bfseries 6}
  (1932)  249.

\bibitem{Kuijken:1989_a}
K.~{Kuijken} and G.~{Gilmore}, ``{The mass distribution in the galactic disc. I
  - A technique to determine the integral surface mass density of the disc near
  the sun.},'' \href{http://dx.doi.org/10.1093/mnras/239.2.571}{{\em \mnras}
  {\bfseries 239} (1989)  571--603}.

\bibitem{Kuijken:1989hu}
K.~Kuijken and G.~Gilmore, ``{The Mass Distribution in the Galactic Disc - Part
  Two - Determination of the Surface Mass Density of the Galactic Disc Near the
  Sun},'' \href{http://dx.doi.org/10.1093/mnras/239.2.605}{{\em \mnras}
  {\bfseries 239} (1989)  605--649}.

\bibitem{Kuijken:1989_c}
K.~{Kuijken} and G.~{Gilmore}, ``{The Mass Distribution in the Galactic Disc -
  Part III - the Local Volume Mass Density},''
  \href{http://dx.doi.org/10.1093/mnras/239.2.651}{{\em \mnras} {\bfseries 239}
  (1989)  651--664}.

\bibitem{Kuijken:1991}
K.~{Kuijken} and G.~{Gilmore}, ``{The Galactic Disk Surface Mass Density and
  the Galactic Force K Z at Z = 1.1 Kiloparsecs},''
  \href{http://dx.doi.org/10.1086/185920}{{\em Astrophys. J.} {\bfseries 367}
  (1991)  L9}.

\bibitem{Perryman:1997sa}
M.~Perryman {\em et al.}, ``{The Hipparcos catalogue},'' {\em Astron.
  Astrophys.} {\bfseries 323} (1997)  L49--L52.

\bibitem{Brown:2018dum}
{\bfseries Gaia} Collaboration, A.~G.~A. Brown {\em et al.}, ``{Gaia Data
  Release 2},'' \href{http://dx.doi.org/10.1051/0004-6361/201833051}{{\em
  Astron. Astrophys.} {\bfseries 616} (2018)  A1},
  \href{http://arxiv.org/abs/1804.09365}{{\ttfamily arXiv:1804.09365
  [astro-ph.GA]}}.

\bibitem{McKee:2015hwa}
C.~F. McKee, A.~Parravano, and D.~J. Hollenbach, ``{Stars, Gas, and Dark Matter
  in the Solar Neighborhood},'' \href{http://dx.doi.org/10.1088/0004-637X}{{\em
  Astrophys. J.} {\bfseries 814} (2015)  13},
  \href{http://arxiv.org/abs/1509.05334}{{\ttfamily arXiv:1509.05334
  [astro-ph.GA]}}.

\bibitem{Xia:2015agz}
Q.~Xia {\em et al.}, ``{Determining the local dark matter density with LAMOST
  data},'' \href{http://dx.doi.org/10.1093/mnras/stw565}{{\em \mnras}
  {\bfseries 458} (2016)  3839},
  \href{http://arxiv.org/abs/1510.06810}{{\ttfamily arXiv:1510.06810
  [astro-ph.GA]}}.

\bibitem{Budenbender:2014xra}
A.~B{\"u}denbender, G.~van~de Ven, and L.~L. Watkins, ``{The tilt of the
  velocity ellipsoid in the Milky Way disc},''
  \href{http://dx.doi.org/10.1093/mnras/stv1314}{{\em \mnras} {\bfseries 452}
  (2015)  956--968}, \href{http://arxiv.org/abs/1407.4808}{{\ttfamily
  arXiv:1407.4808 [astro-ph.GA]}}.

\bibitem{Sivertsson:2017rkp}
S.~Sivertsson, H.~Silverwood, J.~I. Read, G.~Bertone, and P.~Steger, ``{The
  local dark matter density from SDSS-SEGUE G-dwarfs},''
  \href{http://dx.doi.org/10.1093/mnras/sty977}{{\em \mnras} {\bfseries 478}
  (2018)  1677}, \href{http://arxiv.org/abs/1708.07836}{{\ttfamily
  arXiv:1708.07836 [astro-ph.GA]}}.

\bibitem{Hagen:1802.09291}
J.~H.~J. {Hagen} and A.~{Helmi}, ``{The vertical force in the solar
  neighbourhood using red clump stars in TGAS and RAVE. Constraints on the
  local dark matter density},''
  \href{http://dx.doi.org/10.1051/0004-6361/201832903}{{\em Astron. Astrophys.}
  {\bfseries 615} (2018)  A99},
  \href{http://arxiv.org/abs/1802.09291}{{\ttfamily arXiv:1802.09291
  [astro-ph.GA]}}.

\bibitem{Schutz:2017tfp}
K.~Schutz, T.~Lin, B.~R. Safdi, and C.-L. Wu, ``{Constraining a Thin Dark
  Matter Disk with Gaia},''
  \href{http://dx.doi.org/10.1103/PhysRevLett.121.081101}{{\em Phys. Rev.
  Lett.} {\bfseries 121} (2018)  081101},
  \href{http://arxiv.org/abs/1711.03103}{{\ttfamily arXiv:1711.03103
  [astro-ph.GA]}}.

\bibitem{Buch:2018qdr}
J.~Buch, S.~C.~J. Leung, and J.~Fan, ``{Using Gaia DR2 to Constrain Local Dark
  Matter Density and Thin Dark Disk},''
  \href{http://dx.doi.org/10.1088/1475-7516/2019/04/026}{{\em \jcap} {\bfseries
  04} (2019)  026}, \href{http://arxiv.org/abs/1808.05603}{{\ttfamily
  arXiv:1808.05603 [astro-ph.GA]}}.

\bibitem{Widmark:2020vqi}
A.~Widmark, P.~F. de~Salas, and G.~Monari, ``{Weighing the Galactic disk in
  sub-regions of the solar neighbourhood using Gaia DR2},''
  \href{http://arxiv.org/abs/2011.02490}{{\ttfamily arXiv:2011.02490
  [astro-ph.GA]}}.

\bibitem{Bovy:2013raa}
J.~Bovy and H.-W. Rix, ``{A Direct Dynamical Measurement of the Milky Way's
  Disk Surface Density Profile, Disk Scale Length, and Dark Matter Profile at 4
  kpc $\lesssim$ R $\lesssim$ 9 kpc},''
  \href{http://dx.doi.org/10.1088/0004-637X/779/2/115}{{\em Astrophys. J.}
  {\bfseries 779} (2013)  115},
  \href{http://arxiv.org/abs/1309.0809}{{\ttfamily arXiv:1309.0809
  [astro-ph.GA]}}.

\bibitem{Zhang:2012rsb}
L.~Zhang, H.-W. Rix, G.~van~de Ven, J.~Bovy, C.~Liu, and G.~Zhao, ``{The
  Gravitational Potential Near the Sun From SEGUE K-dwarf Kinematics},''
  \href{http://dx.doi.org/10.1088/0004-637X/772/2/108}{{\em Astrophys. J.}
  {\bfseries 772} (2013)  108},
  \href{http://arxiv.org/abs/1209.0256}{{\ttfamily arXiv:1209.0256
  [astro-ph.GA]}}.

\bibitem{Bienayme:2014kva}
O.~Bienaym{\'e} {\em et al.}, ``{Weighing the local dark matter with RAVE red
  clump stars},'' \href{http://dx.doi.org/10.1051/0004-6361/201424478}{{\em
  \aap} {\bfseries 571} (2014)  A92},
  \href{http://arxiv.org/abs/1406.6896}{{\ttfamily arXiv:1406.6896
  [astro-ph.GA]}}.

\bibitem{Silverwood:2015hxa}
H.~Silverwood, S.~Sivertsson, P.~Steger, J.~Read, and G.~Bertone, ``{A
  non-parametric method for measuring the local dark matter density},''
  \href{http://dx.doi.org/10.1093/mnras/stw917}{{\em \mnras} {\bfseries 459}
  (2016)  4191--4208}, \href{http://arxiv.org/abs/1507.08581}{{\ttfamily
  arXiv:1507.08581 [astro-ph.GA]}}.

\bibitem{Zheng:2001wc}
Z.~Zheng, C.~Flynn, A.~Gould, J.~N. Bahcall, and S.~Salim, ``{M dwarfs from
  Hubble Space Telescope star counts. 4},''
  \href{http://dx.doi.org/10.1086/321485}{{\em Astrophys. J.} {\bfseries 555}
  (2001)  393--404}, \href{http://arxiv.org/abs/astro-ph/0102442}{{\ttfamily
  arXiv:astro-ph/0102442}}.

\bibitem{Flynn:2006tm}
C.~Flynn, J.~Holmberg, L.~Portinari, B.~Fuchs, and H.~Jahreiss, ``{On the
  mass-to-light ratio of the local Galactic disc and the optical luminosity of
  the Galaxy},'' \href{http://dx.doi.org/10.1111/j.1365-2966.2006.10911.x}{{\em
  \mnras} {\bfseries 372} (2006)  1149--1160},
  \href{http://arxiv.org/abs/astro-ph/0608193}{{\ttfamily
  arXiv:astro-ph/0608193}}.

\bibitem{Kramer:2016dew}
E.~D. Kramer and L.~Randall, ``{Interstellar Gas and a Dark Disk},''
  \href{http://dx.doi.org/10.3847/0004-637X/829/2/126}{{\em Astrophys. J.}
  {\bfseries 829} (2016)  126},
  \href{http://arxiv.org/abs/1603.03058}{{\ttfamily arXiv:1603.03058
  [astro-ph.GA]}}.

\bibitem{Widmark:1711.07504}
A.~{Widmark} and G.~{Monari}, ``{The dynamical matter density in the solar
  neighbourhood inferred from Gaia DR1},''
  \href{http://dx.doi.org/10.1093/mnras/sty2400}{{\em \mnras} {\bfseries 482}
  (2019)  262--277}, \href{http://arxiv.org/abs/1711.07504}{{\ttfamily
  arXiv:1711.07504 [astro-ph.GA]}}.

\bibitem{Widmark:2018ylf}
A.~Widmark, ``{Measuring the local matter density using Gaia DR2},''
  \href{http://dx.doi.org/10.1051/0004-6361/201834718}{{\em Astron. Astrophys.}
  {\bfseries 623} (2019)  A30},
  \href{http://arxiv.org/abs/1811.07911}{{\ttfamily arXiv:1811.07911
  [astro-ph.GA]}}.

\bibitem{Mroz:1810.02131}
P.~{Mr{\'o}z}, A.~{Udalski}, D.~M. {Skowron}, J.~{Skowron}, I.~{Soszy{\'n}ski},
  P.~{Pietrukowicz}, M.~K. {Szyma{\'n}ski}, R.~{Poleski}, S.~{Koz{\l}owski},
  and K.~{Ulaczyk}, ``{Rotation Curve of the Milky Way from Classical
  Cepheids},'' \href{http://dx.doi.org/10.3847/2041-8213/aaf73f}{{\em
  Astrophys. J. Lett.} {\bfseries 870} (2019)  L10},
  \href{http://arxiv.org/abs/1810.02131}{{\ttfamily arXiv:1810.02131
  [astro-ph.GA]}}.

\bibitem{Pato:2015dua}
M.~Pato, F.~Iocco, and G.~Bertone, ``{Dynamical constraints on the dark matter
  distribution in the Milky Way},''
  \href{http://dx.doi.org/10.1088/1475-7516/2015/12/001}{{\em \jcap} {\bfseries
  12} (2015)  001}, \href{http://arxiv.org/abs/1504.06324}{{\ttfamily
  arXiv:1504.06324 [astro-ph.GA]}}.

\bibitem{Benito:2019ngh}
M.~Benito, A.~Cuoco, and F.~Iocco, ``{Handling the Uncertainties in the
  Galactic Dark Matter Distribution for Particle Dark Matter Searches},''
  \href{http://dx.doi.org/10.1088/1475-7516/2019/03/033}{{\em \jcap} {\bfseries
  03} (2019)  033}, \href{http://arxiv.org/abs/1901.02460}{{\ttfamily
  arXiv:1901.02460 [astro-ph.GA]}}.

\bibitem{Benito:2020lgu}
M.~Benito, F.~Iocco, and A.~Cuoco, ``{Uncertainties in the Galactic dark matter
  distribution: an update},'' \href{http://arxiv.org/abs/2009.13523}{{\ttfamily
  arXiv:2009.13523 [astro-ph.GA]}}.

\bibitem{Karukes:2019jxv}
E.~V. Karukes, M.~Benito, F.~Iocco, R.~Trotta, and A.~Geringer-Sameth,
  ``{Bayesian reconstruction of the Milky Way dark matter distribution},''
  \href{http://dx.doi.org/10.1088/1475-7516/2019/09/046}{{\em JCAP} {\bfseries
  1909} (2019)  046}, \href{http://arxiv.org/abs/1901.02463}{{\ttfamily
  arXiv:1901.02463 [astro-ph.GA]}}.

\bibitem{Huang:1604.01216}
Y.~{Huang}, X.~W. {Liu}, H.~B. {Yuan}, M.~S. {Xiang}, H.~W. {Zhang}, B.~Q.
  {Chen}, J.~J. {Ren}, C.~{Wang}, Y.~{Zhang}, Y.~H. {Hou}, Y.~F. {Wang}, and
  Z.~H. {Cao}, ``{The Milky Way's rotation curve out to 100 kpc and its
  constraint on the Galactic mass distribution},''
  \href{http://dx.doi.org/10.1093/mnras/stw2096}{{\em \mnras} {\bfseries 463}
  (2016)  2623--2639}, \href{http://arxiv.org/abs/1604.01216}{{\ttfamily
  arXiv:1604.01216 [astro-ph.GA]}}.

\bibitem{Lin:2019yux}
H.-N. Lin and X.~Li, ``{The Dark Matter Profiles in the Milky Way},''
  \href{http://dx.doi.org/10.1093/mnras/stz1698}{{\em \mnras} {\bfseries 487}
  (2019)  5679--5684}, \href{http://arxiv.org/abs/1906.08419}{{\ttfamily
  arXiv:1906.08419 [astro-ph.GA]}}.

\bibitem{deSalas:2019pee}
P.~F. de~Salas, K.~Malhan, K.~Freese, K.~Hattori, and M.~Valluri, ``{On the
  estimation of the Local Dark Matter Density using the rotation curve of the
  Milky Way},'' \href{http://dx.doi.org/10.1088/1475-7516/2019/10/037}{{\em
  JCAP} {\bfseries 10} (2019)  037},
  \href{http://arxiv.org/abs/1906.06133}{{\ttfamily arXiv:1906.06133
  [astro-ph.GA]}}.

\bibitem{Chemin:1504.01507}
L.~{Chemin}, F.~{Renaud}, and C.~{Soubiran}, ``{Incorrect rotation curve of the
  Milky Way},'' \href{http://dx.doi.org/10.1051/0004-6361/201526040}{{\em
  Astron. Astrophys.} {\bfseries 578} (2015)  A14},
  \href{http://arxiv.org/abs/1504.01507}{{\ttfamily arXiv:1504.01507
  [astro-ph.GA]}}.

\bibitem{Pato:2017yai}
M.~Pato and F.~Iocco, ``{$\texttt{galkin}$: a new compilation of the Milky Way
  rotation curve data},'' \href{http://arxiv.org/abs/1703.00020}{{\ttfamily
  arXiv:1703.00020 [astro-ph.GA]}}.

\bibitem{Malhotra:1994qj}
S.~Malhotra, ``{The vertical distribution and kinematics of hi and mass models
  of the galactic disk},'' \href{http://dx.doi.org/10.1086/175946}{{\em
  Astrophys. J.} {\bfseries 448} (1995)  138--148}.

\bibitem{Fich:1989}
M.~{Fich}, L.~{Blitz}, and A.~A. {Stark}, ``{The Rotation Curve of the Milky
  Way to 2R 0},'' \href{http://dx.doi.org/10.1086/167591}{{\em Astrophys. J.}
  {\bfseries 342} (1989)  272}.

\bibitem{Hogg:1810.09468}
D.~W. {Hogg}, A.-C. {Eilers}, and H.-W. {Rix}, ``{Spectrophotometric Parallaxes
  with Linear Models: Accurate Distances for Luminous Red-giant Stars},''
  \href{http://dx.doi.org/10.3847/1538-3881/ab398c}{{\em Astronom. J.}
  {\bfseries 158} (2019)  147},
  \href{http://arxiv.org/abs/1810.09468}{{\ttfamily arXiv:1810.09468
  [astro-ph.GA]}}.

\bibitem{Pouliasis:1611.07979}
E.~Pouliasis, P.~D. Matteo, and M.~Haywood, ``{A Milky Way with a massive,
  centrally concentrated thick disc: new Galactic mass models for orbit
  computations},'' \href{http://arxiv.org/abs/1611.07979}{{\ttfamily
  arXiv:1611.07979 [astro-ph.GA]}}.

\bibitem{Misiriotis:2006qq}
A.~Misiriotis, E.~M. Xilouris, J.~Papamastorakis, P.~Boumis, and C.~D. Goudis,
  ``{The distribution of the ISM in the Milky Way A three-dimensional
  large-scale model},''
  \href{http://dx.doi.org/10.1051/0004-6361:20054618}{{\em Astron. Astrophys.}
  {\bfseries 459} (2006)  113}.

\bibitem{Piffl:2015xua}
T.~Piffl, Z.~Penoyre, and J.~Binney, ``{Bringing the Galaxy's dark halo to
  life},'' \href{http://dx.doi.org/10.1093/mnras/stv938}{{\em \mnras}
  {\bfseries 451} (2015)  639--650},
  \href{http://arxiv.org/abs/1502.02916}{{\ttfamily arXiv:1502.02916
  [astro-ph.GA]}}.

\bibitem{Piffl:2014mfa}
T.~Piffl {\em et al.}, ``{Constraining the Galaxy's dark halo with RAVE
  stars},'' \href{http://dx.doi.org/10.1093/mnras/stu1948}{{\em \mnras}
  {\bfseries 445} (2014)  3133},
  \href{http://arxiv.org/abs/1406.4130}{{\ttfamily arXiv:1406.4130
  [astro-ph.GA]}}.

\bibitem{Binney:2015gaa}
J.~Binney and T.~Piffl, ``{The distribution function of the Galaxy's dark
  halo},'' \href{http://dx.doi.org/10.1093/mnras/stv2225}{{\em \mnras}
  {\bfseries 454} (2015)  3653},
  \href{http://arxiv.org/abs/1509.06877}{{\ttfamily arXiv:1509.06877
  [astro-ph.GA]}}.

\bibitem{Cole:2016gzv}
D.~R. Cole and J.~Binney, ``{A centrally heated dark halo for our Galaxy},''
  \href{http://dx.doi.org/10.1093/mnras/stw2775}{{\em \mnras} {\bfseries 465}
  (2017)  798}, \href{http://arxiv.org/abs/1610.07818}{{\ttfamily
  arXiv:1610.07818 [astro-ph.GA]}}.

\bibitem{Cautun:2019eaf}
M.~Cautun {\em et al.}, ``{The Milky Way total mass profile as inferred from
  Gaia DR2},'' \href{http://dx.doi.org/10.1093/mnras/staa1017}{{\em \mnras}
  {\bfseries 494} (2020)  4291--4313},
  \href{http://arxiv.org/abs/1911.04557}{{\ttfamily arXiv:1911.04557
  [astro-ph.GA]}}.

\bibitem{Juric:2005zr}
{\bfseries SDSS} Collaboration, M.~Juric {\em et al.}, ``{The Milky Way
  Tomography with SDSS. 1. Stellar Number Density Distribution},''
  \href{http://dx.doi.org/10.1086/523619}{{\em Astrophys. J.} {\bfseries 673}
  (2008)  864--914}.

\bibitem{Cappellari:1907.09894}
M.~{Cappellari}, ``{Efficient solution of the anisotropic spherically aligned
  axisymmetric Jeans equations of stellar hydrodynamics for galactic
  dynamics},'' \href{http://dx.doi.org/10.1093/mnras/staa959}{{\em \mnras}
  {\bfseries 494} (2020)  4819--4837},
  \href{http://arxiv.org/abs/1907.09894}{{\ttfamily arXiv:1907.09894
  [astro-ph.GA]}}.

\bibitem{Callingham:1808.10456}
T.~M. {Callingham}, M.~{Cautun}, A.~J. {Deason}, C.~S. {Frenk}, W.~{Wang},
  F.~A. {G{\'o}mez}, R.~J.~J. {Grand}, F.~{Marinacci}, and R.~{Pakmor}, ``{The
  mass of the Milky Way from satellite dynamics},''
  \href{http://dx.doi.org/10.1093/mnras/stz365}{{\em \mnras} {\bfseries 484}
  (2019)  5453--5467}, \href{http://arxiv.org/abs/1808.10456}{{\ttfamily
  arXiv:1808.10456 [astro-ph.GA]}}.

\bibitem{Sesar:1611.08596}
B.~{Sesar}, N.~{Hernitschek}, S.~{Mitrovi{\'c}}, {\v{Z}}.~{Ivezi{\'c}}, H.-W.
  {Rix}, J.~G. {Cohen}, E.~J. {Bernard}, E.~K. {Grebel}, N.~F. {Martin}, E.~F.
  {Schlafly}, W.~S. {Burgett}, P.~W. {Draper}, H.~{Flewelling}, N.~{Kaiser},
  R.~P. {Kudritzki}, E.~A. {Magnier}, N.~{Metcalfe}, J.~L. {Tonry}, and
  C.~{Waters}, ``{Machine-learned Identification of RR Lyrae Stars from Sparse,
  Multi-band Data: The PS1 Sample},''
  \href{http://dx.doi.org/10.3847/1538-3881/aa661b}{{\em Astron. J.} {\bfseries
  153} (2017)  204}, \href{http://arxiv.org/abs/1611.08596}{{\ttfamily
  arXiv:1611.08596 [astro-ph.GA]}}.

\bibitem{Portail:1502.00633}
M.~{Portail}, C.~{Wegg}, O.~{Gerhard}, and I.~{Martinez-Valpuesta},
  ``{Made-to-measure models of the Galactic box/peanut bulge: stellar and total
  mass in the bulge region},''
  \href{http://dx.doi.org/10.1093/mnras/stv058}{{\em \mnras} {\bfseries 448}
  (2015)  713--731}, \href{http://arxiv.org/abs/1502.00633}{{\ttfamily
  arXiv:1502.00633 [astro-ph.GA]}}.

\bibitem{Iorio:1808.04370}
G.~{Iorio} and V.~{Belokurov}, ``{The shape of the Galactic halo with Gaia DR2
  RR Lyrae. Anatomy of an ancient major merger},''
  \href{http://dx.doi.org/10.1093/mnras/sty2806}{{\em \mnras} {\bfseries 482}
  (2019)  3868--3879}, \href{http://arxiv.org/abs/1808.04370}{{\ttfamily
  arXiv:1808.04370 [astro-ph.GA]}}.

\bibitem{Laporte:1808.00451}
C.~F.~P. Laporte, I.~Minchev, K.~V. Johnston, and F.~A. G{\'o}mez,
  ``{Footprints of the Sagittarius dwarf galaxy in the $Gaia$ data set},''
  \href{http://dx.doi.org/10.1093/mnras/stz583}{{\em \mnras} {\bfseries 485}
  (2019)  3134}, \href{http://arxiv.org/abs/1808.00451}{{\ttfamily
  arXiv:1808.00451 [astro-ph.GA]}}.

\bibitem{Haines:1903.00607}
T.~{Haines}, E.~{D'Onghia}, B.~{Famaey}, C.~{Laporte}, and L.~{Hernquist},
  ``{Implications of a Time-varying Galactic Potential for Determinations of
  the Dynamical Surface Density},''
  \href{http://dx.doi.org/10.3847/2041-8213/ab25f3}{{\em \apjl} {\bfseries 879}
  (2019)  L15}, \href{http://arxiv.org/abs/1903.00607}{{\ttfamily
  arXiv:1903.00607 [astro-ph.GA]}}.

\bibitem{Banik:2016yqm}
N.~Banik, L.~M. Widrow, and S.~Dodelson, ``{Galactoseismology and the Local
  Density of Dark Matter},''
  \href{http://dx.doi.org/10.1093/mnras/stw2603}{{\em \mnras} {\bfseries 464}
  (2017)  3775}, \href{http://arxiv.org/abs/1608.03338}{{\ttfamily
  arXiv:1608.03338 [astro-ph.GA]}}.

\bibitem{Yanny:2013pqi}
B.~Yanny and S.~Gardner, ``{The Stellar Number Density Distribution in the
  Local Solar Neighborhood is North-South Asymmetric},''
  \href{http://dx.doi.org/10.1088/0004-637X/777/2/91}{{\em Astrophys. J.}
  {\bfseries 777} (2013)  91}, \href{http://arxiv.org/abs/1309.2300}{{\ttfamily
  arXiv:1309.2300 [astro-ph.GA]}}.

\bibitem{Bennett:1809.03507}
M.~{Bennett} and J.~{Bovy}, ``{Vertical waves in the solar neighbourhood in
  Gaia DR2},'' \href{http://dx.doi.org/10.1093/mnras/sty2813}{{\em \mnras}
  {\bfseries 482} (2019)  1417--1425},
  \href{http://arxiv.org/abs/1809.03507}{{\ttfamily arXiv:1809.03507
  [astro-ph.GA]}}.

\bibitem{Widrow:2012wu}
L.~M. Widrow, S.~Gardner, B.~Yanny, S.~Dodelson, and H.-Y. Chen,
  ``{Galactoseismology: Discovery of Vertical Waves in the Galactic Disk},''
  \href{http://dx.doi.org/10.1088/2041-8205/750/2/L41}{{\em Astrophys. J.
  Lett.} {\bfseries 750} (2012)  L41},
  \href{http://arxiv.org/abs/1203.6861}{{\ttfamily arXiv:1203.6861
  [astro-ph.GA]}}.

\bibitem{Gomez:2012rd}
F.~A. Gomez, I.~Minchev, B.~W. O'Shea, T.~C. Beers, J.~S. Bullock, and C.~W.
  Purcell, ``{Vertical density waves in the Milky Way disc induced by the
  Sagittarius Dwarf Galaxy},''
  \href{http://dx.doi.org/10.1093/mnras/sts327}{{\em \mnras} {\bfseries 429}
  (2013)  159}, \href{http://arxiv.org/abs/1207.3083}{{\ttfamily
  arXiv:1207.3083 [astro-ph.GA]}}.

\bibitem{Williams:2013pbz}
M.~Williams {\em et al.}, ``{The wobbly Galaxy: kinematics north and south with
  RAVE red clump giants},'' \href{http://dx.doi.org/10.1093/mnras/stt1522}{{\em
  \mnras} {\bfseries 436} (2013)  101},
  \href{http://arxiv.org/abs/1302.2468}{{\ttfamily arXiv:1302.2468
  [astro-ph.GA]}}.

\bibitem{Carlin:2013eba}
J.~L. Carlin {\em et al.}, ``{Substructure in bulk velocities of Milky Way disk
  stars},'' \href{http://dx.doi.org/10.1088/2041-8205/777/1/L5}{{\em Astrophys.
  J. Lett.} {\bfseries 777} (2013)  L5},
  \href{http://arxiv.org/abs/1309.6314}{{\ttfamily arXiv:1309.6314
  [astro-ph.GA]}}.

\bibitem{Widrow:2014jxa}
L.~M. Widrow, J.~Barber, M.~H. Chequers, and E.~Cheng, ``{Bending and breathing
  modes of the Galactic disc},''
  \href{http://dx.doi.org/10.1093/mnras/stu396}{{\em \mnras} {\bfseries 440}
  (2014)  1971--1981}, \href{http://arxiv.org/abs/1404.4069}{{\ttfamily
  arXiv:1404.4069 [astro-ph.GA]}}.

\bibitem{Hunt:1806.02832}
J.~A.~S. {Hunt}, J.~{Hong}, J.~{Bovy}, D.~{Kawata}, and R.~J.~J. {Grand},
  ``{Transient spiral structure and the disc velocity substructure in Gaia
  DR2},'' \href{http://dx.doi.org/10.1093/mnras/sty2532}{{\em \mnras}
  {\bfseries 481} (2018)  3794--3803},
  \href{http://arxiv.org/abs/1806.02832}{{\ttfamily arXiv:1806.02832
  [astro-ph.GA]}}.

\bibitem{Lopez-Corredoira:2001.05455}
M.~{L{\'o}pez-Corredoira}, F.~{Garz{\'o}n}, H.~F. {Wang}, F.~{Sylos Labini},
  R.~{Nagy}, {\v{Z}}.~{Chrob{\'a}kov{\'a}}, J.~{Chang}, and B.~{Villarroel},
  ``{Gaia-DR2 extended kinematical maps. II. Dynamics in the Galactic disk
  explaining radial and vertical velocities},''
  \href{http://dx.doi.org/10.1051/0004-6361/201936711}{{\em Astron. Astrophys.}
  {\bfseries 634} (2020)  A66},
  \href{http://arxiv.org/abs/2001.05455}{{\ttfamily arXiv:2001.05455
  [astro-ph.GA]}}.

\bibitem{Monari:1606.06785}
G.~{Monari}, B.~{Famaey}, A.~{Siebert}, R.~J.~J. {Grand }, D.~{Kawata}, and
  C.~{Boily}, ``{The effects of bar-spiral coupling on stellar kinematics in
  the Galaxy},'' \href{http://dx.doi.org/10.1093/mnras/stw1564}{{\em \mnras}
  {\bfseries 461} (2016)  3835--3846},
  \href{http://arxiv.org/abs/1606.06785}{{\ttfamily arXiv:1606.06785
  [astro-ph.GA]}}.

\bibitem{Michtchenko:1608.08991}
T.~A. {Michtchenko}, R.~S.~S. {Vieira}, D.~A. {Barros}, and J.~R.~D.
  {L{\'e}pine}, ``{Modelling resonances and orbital chaos in disk galaxies.
  Application to a Milky Way spiral model},''
  \href{http://dx.doi.org/10.1051/0004-6361/201628895}{{\em Astron. Astrophys.}
  {\bfseries 597} (2017)  A39},
  \href{http://arxiv.org/abs/1608.08991}{{\ttfamily arXiv:1608.08991
  [astro-ph.GA]}}.

\bibitem{Vasiliev:2009.10726}
E.~{Vasiliev}, V.~{Belokurov}, and D.~{Erkal}, ``{Tango for three: Sagittarius,
  LMC, and the Milky Way},''
  \href{http://dx.doi.org/10.1093/mnras/staa3673}{{\em \mnras} (2020)  },
  \href{http://arxiv.org/abs/2009.10726}{{\ttfamily arXiv:2009.10726
  [astro-ph.GA]}}.

\bibitem{Hinkel:2020tii}
A.~Hinkel, S.~Gardner, and B.~Yanny, ``{Probing Axial Symmetry Breaking in the
  Galaxy with Gaia Data Release 2},''
  \href{http://dx.doi.org/10.3847/1538-4357/ab8235}{{\em Astrophys. J.}
  {\bfseries 893} (2020)  105},
  \href{http://arxiv.org/abs/2003.08389}{{\ttfamily arXiv:2003.08389
  [astro-ph.GA]}}.

\bibitem{Bovy:1610.07610}
J.~{Bovy}, ``{Galactic rotation in Gaia DR1},''
  \href{http://dx.doi.org/10.1093/mnrasl/slx027}{{\em \mnras} {\bfseries 468}
  (2017)  L63--L67}, \href{http://arxiv.org/abs/1610.07610}{{\ttfamily
  arXiv:1610.07610 [astro-ph.GA]}}.

\bibitem{Hessman:2015nua}
F.~V. Hessman, ``{The difficulty of measuring the local dark matter density},''
  \href{http://dx.doi.org/10.1051/0004-6361/201526022}{{\em Astron. Astrophys.}
  {\bfseries 579} (2015)  A123},
  \href{http://arxiv.org/abs/1506.00384}{{\ttfamily arXiv:1506.00384
  [astro-ph.GA]}}.

\bibitem{Lallement:2003_a}
R.~{Lallement}, B.~Y. {Welsh}, J.~L. {Vergely}, F.~{Crifo}, and D.~{Sfeir},
  ``{3D mapping of the dense interstellar gas around the Local Bubble},''
  \href{http://dx.doi.org/10.1051/0004-6361:20031214}{{\em Astron. Astrophys.}
  {\bfseries 411} (2003)  447--464}.

\bibitem{Salucci:2010qr}
P.~Salucci, F.~Nesti, G.~Gentile, and C.~Martins, ``{The dark matter density at
  the Sun's location},''
  \href{http://dx.doi.org/10.1051/0004-6361/201014385}{{\em Astron. Astrophys.}
  {\bfseries 523} (2010)  A83},
  \href{http://arxiv.org/abs/1003.3101}{{\ttfamily arXiv:1003.3101
  [astro-ph.GA]}}.

\bibitem{Lake:1989.AJ.98.1554}
G.~{Lake}, ``{Must the Disk and Halo Dark Matter Be Different?},''
  \href{http://dx.doi.org/10.1086/115238}{{\em Astron. J.} {\bfseries 98}
  (1989)  1554}.

\bibitem{OHare:2018trr}
C.~A. O'Hare, C.~McCabe, N.~W. Evans, G.~Myeong, and V.~Belokurov, ``{Dark
  matter hurricane: Measuring the S1 stream with dark matter detectors},''
  \href{http://dx.doi.org/10.1103/PhysRevD.98.103006}{{\em Phys. Rev. D}
  {\bfseries 98} (2018)  103006},
  \href{http://arxiv.org/abs/1807.09004}{{\ttfamily arXiv:1807.09004
  [astro-ph.CO]}}.

\bibitem{Myeong:2017skt}
G.~Myeong, N.~Evans, V.~Belokurov, N.~Amorisco, and S.~Koposov, ``{Halo
  Substructure in the SDSS-Gaia Catalogue : Streams and Clumps},''
  \href{http://dx.doi.org/10.1093/mnras/stx3262}{{\em \mnras} {\bfseries 475}
  (2018)  1537--1548}, \href{http://arxiv.org/abs/1712.04071}{{\ttfamily
  arXiv:1712.04071 [astro-ph.GA]}}.

\bibitem{Klioner:2012.02036}
{\bfseries Gaia} Collaboration, S.~A. Klioner {\em et al.}, ``{Gaia Early Data
  Release 3: Acceleration of the solar system from Gaia astrometry},''
  \href{http://arxiv.org/abs/2012.02036}{{\ttfamily arXiv:2012.02036
  [astro-ph.GA]}}.

\bibitem{Bovy:2012.02169}
J.~Bovy, ``{A purely acceleration-based measurement of the fundamental Galactic
  parameters using Gaia EDR3},''
  \href{http://arxiv.org/abs/2012.02169}{{\ttfamily arXiv:2012.02169
  [astro-ph.GA]}}.

\bibitem{Ravi:2018vqd}
A.~Ravi, N.~Langellier, D.~F. Phillips, {\em et al.}, ``{Probing Dark Matter
  Using Precision Measurements of Stellar Accelerations},''
  \href{http://dx.doi.org/10.1103/PhysRevLett.123.091101}{{\em Phys. Rev.
  Lett.} {\bfseries 123} (2019)  091101},
  \href{http://arxiv.org/abs/1812.07578}{{\ttfamily arXiv:1812.07578
  [astro-ph.GA]}}.

\bibitem{Silverwood:2018qra}
H.~Silverwood and R.~Easther, ``{Stellar Accelerations and the Galactic
  Gravitational Field},'' \href{http://dx.doi.org/10.1017/pasa.2019.25}{{\em
  Publ. Astron. Soc. Austral.} {\bfseries 36} (2019)  e038},
  \href{http://arxiv.org/abs/1812.07581}{{\ttfamily arXiv:1812.07581
  [astro-ph.GA]}}.

\bibitem{Widmark:2020zad}
A.~Widmark, K.~Malhan, P.~F. de~Salas, and S.~Sivertsson, ``{Measuring the
  matter density of the Galactic disc using stellar streams},''
  \href{http://dx.doi.org/10.1093/mnras/staa1741}{{\em \mnras} {\bfseries 496}
  (2020)  3112--3127}, \href{http://arxiv.org/abs/2003.04318}{{\ttfamily
  arXiv:2003.04318 [astro-ph.GA]}}.

\bibitem{Malhan:2018_streamfinder}
K.~Malhan and R.~A. Ibata, ``{STREAMFINDER -- I. A new algorithm for detecting
  stellar streams},'' \href{http://dx.doi.org/10.1093/mnras/sty912}{{\em
  \mnras} {\bfseries 477} (2018)  4063--4076}.

\bibitem{Borsato:1907.02527}
N.~W. {Borsato}, S.~L. {Martell}, and J.~D. {Simpson}, ``{Identifying stellar
  streams in Gaia DR2 with data mining techniques},''
  \href{http://dx.doi.org/10.1093/mnras/stz3479}{{\em \mnras} {\bfseries 492}
  (2020)  1370--1384}, \href{http://arxiv.org/abs/1907.02527}{{\ttfamily
  arXiv:1907.02527 [astro-ph.GA]}}.

\bibitem{Kavanagh:2020cvn}
B.~J. Kavanagh, T.~Emken, and R.~Catena, ``{Measuring the local Dark Matter
  density in the laboratory},''
  \href{http://arxiv.org/abs/2004.01621}{{\ttfamily arXiv:2004.01621
  [hep-ph]}}.

\bibitem{Berge:2019zjj}
J.~Berg{\'e} {\em et al.}, ``{The local dark sector. Probing gravitation's
  low-acceleration frontier and dark matter in the Solar System
  neighborhood},'' \href{http://arxiv.org/abs/1909.00834}{{\ttfamily
  arXiv:1909.00834 [astro-ph.IM]}}.

\end{thebibliography}

\end{document}